\documentstyle[11pt,psfig,aaspp4]{article}
\begin{document}

\def \eff {{\rm eff}}
\def \FS {{\it FS\/}}
\def \MB {{\it MB\/}}
\def \HPBW {{\it HPBW\/}}
\def \dKperJy {$K\!p\!e\!r\!J\!y$}
\def \KperJy {K\!p\!e\!r\!J\!y}
\def \az {{\it az}}
\def \za {{\it za}}
\def \coma {{\rm coma}}
\def \beam {{\rm beam}}

\title{ALL-STOKES PARAMETERIZATION OF THE MAIN BEAM AND FIRST SIDELOBE 
FOR THE ARECIBO RADIO TELESCOPE}


\author{Carl Heiles}

\affil {Astronomy Department, University of California,
    Berkeley, CA 94720-3411; cheiles@astron.berkeley.edu}

\author{Phil Perillat, Michael Nolan, Duncan Lorimer, Ramesh Bhat, 
Tapasi Ghosh, Ellen Howell, Murray Lewis, Karen O'Neil, Chris Salter, 
Snezana Stanimirovic}
\affil {Arecibo Observatory, Arecibo, PR 00613; email addresses
available at {\it www.naic.edu}}

\begin{abstract}
   
	Radio astronomical measurements of extended emission require
knowledge of the beam shape and response because the measurements need
correction for quantities such as beam efficiency and beamwidth.  We
describe a scheme that characterizes the main beam and sidelobe in all
Stokes parameters employing parameters that allow reconstruction of the
complete beam patterns and, also, afford an easy way to see how the beam
changes with azimuth, zenith angle, and time.  For the main beam in
Stokes $I$ the parameters include the beam width, ellipticity and its
orientation, coma and its orientation, the point-source gain, the
integrated gain (or, equivalently, the main beam efficiency); for the
other Stokes parameters the beam parameters include beam squint and beam
squash.  For the first sidelobe ring in Stokes $I$ the parameters
include an 8-term Fourier series describing the height, radius, and
radial width; for the other Stokes parameters they include only the
sidelobe's fractional polarization.  

	We illustrate the technique by applying it to the Arecibo
telescope. The main beam width is smaller and the sidelobe levels higher
than for a uniformly-illuminated aperture of the same effective area. 
These effects are modeled modestly well by a blocked aperture, with the
blocked area equal to about $10\%$ of the effective area (this
corresponds to $5\%$ physical blockage).  In polarized emission, the
effects of beam squint (difference in pointing direction between
orthogonal polarizations) and squash (difference in beamwidth between
orthogonal polarizations) do not correspond to theoretical expectation
and are higher than expected; these effects are almost certainly caused
by the blockage.  The first sidelobe is highly polarized because of
blockage.  These polarization effects lead to severe contamination of
maps of polarized emission by spatial derivatives in brightness
temperature.

\end{abstract}


\section{INTRODUCTION}  \label{introduction}

	The Arecibo\footnote{The Arecibo Observatory is part of the
National Astronomy and Ionosphere Center, which is operated by Cornell
University under a cooperative agreement with the National Science
Foundation} radio telescope is unique because the feed moves
with respect to a reflector that is fixed with respect to the ground. 
Thus, the properties of the telescope beam can change in a complicated
way as a source is tracked because some components of the beam rotate
with the parallactic angle and some don't.  For example, the beam is
significantly elliptical, and the ellipse rotates with parallactic
angle; in contrast, the effects of reflector surface imperfections are
fixed on the ground. Disentangling these complicated behaviors is best
done by parameterizing the beam: some parameters change in recognizable
and physically reasonable ways. 

	Our efforts were directed towards characterizing all of
Arecibo's receivers with the partial intent of establishing a historical
record so that the beam patterns could be monitored as future
improvements in the telescope structure occur.  However, our efforts are
also useful for the process of interpreting observational results, and
can easily be applied to other telescopes.  This paper is a condensed
version of Arecibo Technical and Operations Memo 2000-03, which covers
all of Arecibo's receivers and with more detail.  

	We devised a simple and fast observing pattern that provides
nearly complete information on the main beam and first sidelobe in all
four Stokes parameters.  We use the digital crosscorrelation technique
to derive the Stokes parameters (Heiles 2001).  The pattern consists of
four continuously sampled scans oriented every $45^\circ$ with respect
to $(\hat\az,\hat\za)$ yielding an 8-pointed star-shaped pattern that
extends out to three nominal HPBW (Half Power Beamwidths) on each side
of the nominal pointing position. 

	These data allow parameterization of the main beam and first
sidelobe as described in the following sections.  We begin in \S
\ref{mainbeami} by describing the parameterization scheme for Stokes
$I$.  \S\ref{repvals} describes representative Stokes $I$ results for
the main beam parameters.  \S\ref{unifill} compares those results with a
theoretical model of a uniformly-illuminated blocked circular aperture. 
\S\ref{fs} presents Stokes $I$ results on the first sidelobe. 
\S\ref{otherstokes} defines and describes the parameters beam squint and
squash for the polarized Stokes parameters, presents the polarization of
the first sidelobe, and calculates the levels of contamination caused by
spatial variations in brightness temperature on maps of polarized
emission. \S \ref{greyfigs} presents greyscale images of the beam in all
four Stokes parameters.

\section{ PARAMETERIZATION OF THE STOKES I BEAM} \label{mainbeami}

\subsection{ Five Parameters that Describe the Main Beam Shape}

        We follow Rohlfs and Wilson's (2000) definition of the 
normalized main-beam power pattern, $P(\theta, \phi)_n = {P(\theta,
\phi) \over P_{\it max}}$. We represent it with five parameters of a
two-dimensional Gaussian defining a power pattern with ellipticity and
coma. The parameters are the average beamwidth $\Theta_0$, beam
ellipticity $\Theta_1$ with an orientation $\phi_\beam$, and coma
$\alpha_\coma$ with orientation $\phi_\coma$. 

	To parameterize the elliptical beam, we take the ratio of major
to minor axes to be $\Theta_0 + \Theta_1 \over \Theta_0 - \Theta_1$ with
the major axis oriented at position angle  $\phi_\beam$, and the
beamwidth $\Theta$ to vary as

\begin{equation} \label{beamellipticity}
\Theta = \Theta_0 + \Theta_1 \cos 2 (\phi - \phi_\beam) \; .
\end{equation}

\noindent where $\phi$ is the position angle in the $(\az,\za)$
coordinate system, defined to be zero along the positive azimuth axis
and increase towards positive zenith angle (see Figure~\ref{ibeameg}).

	We parameterize the coma lobe with a function chosen for
analytical convenience; it is not based on a proper derivation.  The
parameters are the coma strength $\alpha_{\coma}$ and its position angle
$\phi_{\coma}$. We assume that the effect of coma depends only on the
projection of $\theta$ along the coma direction; this distance is 

\begin{equation} 
\theta_{\coma} = \theta \cos(\phi - \phi_{\coma}) \; .
\end{equation}

\noindent The normalized power distribution including parameterized ellipticity
and coma is then:

\begin{equation}
\label{II1}   
P_n = \exp \left[ -
{\theta^2 
\left( 1 - 
\min\{ \alpha_{\coma} {\theta_{\coma} 
\over \Theta_0} , 0.75\} 
\right) \over \Theta^2} 
\right]
\end{equation}

\noindent where $\theta$ is the great-circle angular distance of the 
position from the {\it true} beam center, equal to $\sqrt{\az_{\rm
offset}^2 + \za_{\rm offset}^2}$.  The term $\min\{ \alpha_{\coma}
{\theta_{\coma} \over \Theta_0} , 0.75\}$ is included to prevent the
coma term from unduly distorting the beam far from the center. Note that
the beamwidth $\Theta$ related to the usual half-power beamwidth by
$\HPBW = 2 (\ln 2)^{1/2} \Theta \approx 1.665 \Theta$.

        Figure~\ref{ibeameg} illustrates these quantities with a main
beam having severe coma and exaggerated ellipticity. In this example,
$\alpha_{\coma} = 0.2$, $\phi_{\coma} = 22.5^\circ$, $\Theta_0 = 3.4$
arcmin, $\Theta_1 = 1$ arcmin, and $\phi_\beam = -67.5^\circ$.   

\subsection{Relating the Beam to the Observation Pattern}

When observing the beam shape we must point the telescope to an
assumed beam center. The four scans of the observing pattern intersect
this assumed beam center, as shown by the dash-dot lines on
Figure~\ref{ibeameg}. The difference between the assumed and actual
centers is the pointing offset. Thus, in equation~\ref{II1},
\begin{mathletters}
\begin{equation}
\az_{\rm offset} = \az_{\rm obs} + \delta \az
\end{equation}
\begin{equation}
\za_{\rm offset} = \za_{\rm obs} + \delta \za
\end{equation}
\end{mathletters}
where $(\az_{\rm obs}, \za_{\rm obs})$ are the ``observed'' offsets,
that is the ones measured with respect to the assumed center. 

\subsection{The First Sidelobe in Stokes I} \label{fsintro}

        Detailed maps of Arecibo's first sidelobe show some azimuthal
structure (i.e., with $\phi$ on Figure~\ref{ibeameg}) (Heiles 2000). 
Our observing pattern provides four radial cuts through the beam.  We
least-squares fit each radial cut with three Gaussians, two weak ones at
the beginning and end for the sidelobes and a strong one for the main
beam.  This yields values of the sidelobe height, center, and width at 8
$\phi$'s.  For each one, e.g.\ the height, we Fourier transform  the 8
values and derive 8 complex Fourier coefficients, only 5 of which are
independent because the inputs are real numbers. We reconstruct the
sidelobe using $N$ Fourier coefficients, with $N \le 8$.  Using $N < 8$
is not unreasonable because the measurement and Gaussian fitting
processes are hardly perfect; it is equivalent to performing a
least-squares fit on the 8 $\phi$'s and deriving fewer coefficients.  In
practice, we reconstruct the sidelobe with  $N=6$, which is just
sufficient to reveal the blockage caused by the equilateral triangle
feed support structure, which is a major structural component.

        This procedure is complicated in practice. In some cuts the
sidelobe is absent or is not well-fit. We test each derived sidelobe
Gaussian fit and accept it only if its width is smaller than the nominal
\HPBW and greater than 0.3 times the nominal \HPBW. We also require the
fractional uncertainty in the derived intensity to be less than 0.45. In
cases having no acceptable fit, we take the height equal to zero,
and take the center and width equal to the average of all the other
acceptable centers and widths for that particular group of cuts.

\subsection{Effective Area, On-Axis Gain, and Beam Efficiencies}

        Again we follow Rohlfs and Wilson (2000) and define the
directive gain $G$ as the gain relative to an isotropic radiator for a
single polarization. Here we are concerned with treating polarization,
for which purpose we must explicitly define the {\it unsubscripted}
directive gain to apply to $\rm{Stokes} \  I \over 2$; below we will
subscript it with a polarization label. Conservation of energy
requires that $G$
satisfy the relationship
\begin{equation} \label{gainintegral}
{\int \int}_{\rm whole \ sky} G d\Omega = 4 \pi \; . 
\end{equation}
The on-axis point-source directive gain is directly related to
the effective area $A_{\eff}$ by
\begin{equation}
G_{\rm max} = {4 \pi A_{\eff} \over \lambda^2} \; ,
\end{equation}
and in practical units of source flux in Jy, area in m$^2$ we have
\begin{equation}
\KperJy= 10^{-26} {A_{\eff} \over 2k_B}
\end{equation}
or
\begin{equation}
 A_{\eff} = 2 \times 10^{26} k_B \KperJy
\end{equation} 
so, eliminating $A_{\eff}$
\begin{eqnarray}
G_{\rm max}&=&{4\pi (2 \times 10^{26} k_B \KperJy) \over \lambda^2}\\
&=&3.47 \times 10^8 \KperJy \over \lambda^2_{\rm cm}
\end{eqnarray}

\noindent where $\lambda_{\rm cm}$ is the wavelength in cm. Arecibo's
theoretical gain, without blockage, is $\KperJy \approx 11$, which is
equivalent to a uniformly illuminated circular aperture with radius 94
m. The normalized power pattern $P_n$, which we used above in
equation~\ref{II1}, is just

\begin{equation}
P_n = {G \over G_{\rm max}} = 2.89 \times 10^{-9} {G \lambda_{\rm cm}^2 \over
\KperJy} \; ,
\end{equation}
so 
\begin{equation} 
{\int \int}_{\rm whole \ sky} P_n d\Omega = 3.62 \times 10^{-8}
{\lambda_{\rm cm}^2
\over \KperJy} \; .
\end{equation}

        The ``efficiency'' of a portion of the telescope's beam is
defined as the integral of its directive gain divided its integral over
the whole sky. Thus, using units of arcmin$^2$ for $\Omega$ and cm for
$\lambda$, 

\begin{equation} \label{etabeam}
\eta_\beam = 2.34 {\KperJy \over \lambda_{\rm cm}^2} 
{\int \int}_\beam P_n d\Omega_{\rm arcmin^2} \; .
\end{equation}

\noindent Normally we consider the main beam efficiency, but we can also
use the same terminology for the sidelobes. The main beam and first
sidelobe integrals are, in principle, expressible analytically by our
parameterization scheme; in practice, however, we compute them
numerically. 

\section{REPRESENTATIVE OBSERVED PARAMETER VALUES} \label{repvals}

        Table~\ref{repdata} provides representative observed values for
the above-defined parameters for two receivers at Arecibo Observatory,
measured during September 2000. 
Both are point-source feeds used with the new Gregorian three-mirror
system. The point-source gain at 430 MHz is close to that expected for a
perfect telescope, while in for the L-band Wide receiver (LBW) the gain
decreases with frequency. This is almost certainly a result of
irregularities in the main reflector; a program of adjustments is
planned for early 2001.

\subsection{ Main Beam Ellipticity and Coma }

        Arecibo's beam is elliptical by design; $\phi_\beam$ should be
$90^\circ$, meaning that the $\za$ beam is larger than the $\az$ beam. 
In practice we find that $\phi_{\it beam}$ exhibits systematic
departures from $90^\circ$, ranging from $\sim 80^\circ \rightarrow
110^\circ$ and does not get noticeably worse with frequency. We believe
this is a result of blockage.

	In September 2000 Arecibo had a noticeable coma lobe.  It gets
worse with frequency, with strength $\alpha_{\coma}$ rising from about
0.015 to 0.15 from 430 to 4500 MHz (at $\za \sim 10^\circ$).   The
direction $\phi_{\coma}$ exhibits no clear systematic departure from
being constant as a function of $(az,za)$ or frequency, scattering about
$\phi \sim 25^\circ$.  This suggests that it is tied to the azimuth arm
and therefore probably a result of subreflector misalignment. 

\section{ COMPARISON WITH A THEORETICAL MODEL: A UNIFORMLY-ILLUMINATED
CIRCULAR APERTURE HAVING BLOCKAGE} \label{unifill}

	It is instructive to compare Arecibo's beam pattern with a
theoretical model.  The most convenient is a uniformly-illuminated
circular aperture.  We will then consider a circular aperture with a
central circular blockage.  Even though Arecibo's beam is elliptical,
it's close enough to circular that the model makes physical sense. 

\subsection{ A Uniformly-Illuminated Unblocked Circular Aperture }

	In this section we will define and evaluate the theoretical 
parameters for an unblocked aperture, with which we will compare our
corresponding observed parameters. This comparison leads to a profusion
of subscripts, and to avoid confusion we will first define those
subscripts. The subscript ``$U$'' denotes the theoretical result for a
uniform illumination pattern. There are two forms of such results. One
is obtained directly from the theory beginning with equation
\ref{efield} below, and this is denoted with the additional subscript
``$I$''. The other is an approximate theoretical result obtained by
fitting with Gaussians (as we do the observational data) the
aforementioned direct theoretical result (subscripted with ``$I$'');
this approximation is the one we compare with the data, and has no
additional subscript. In addition, the subscripts ``$MB$'' and  ``$FS$''
refer to the main beam and first sidelobe, respectively. 

The electric field pattern of an unblocked aperture is
\begin{equation}\label{efield}
F(\sin \theta) = 2 {J_1(u)\over u},
\end{equation}
where 
\begin{displaymath}
\begin{array}{ccc}
u = {\sin \theta \over (\theta_d / \pi)}&{\rm and}&\theta_d =
\lambda/D
\end{array}
\end{displaymath}
(Rohlfs and Wilson 2000). The field for a blocked aperture is a
suitably-weighted arithmetic combination of different diameters. In
principle, the best way to fit this model to the data is to compare
the observed and theoretical beam shapes. However, this is not so easy
because of Arecibo's elliptical beam.  Rather, we examine five
parameters. Of these, two are {\it robust} in the sense that they
don't depend on assuming a perfectly smooth reflecting surface for the
bowl and, in addition, don't depend on knowing an accurate source flux
or a noise diode calibration standard (cal) temperature.  

	The robust parameters specify only the beam shape using ratios
of the first sidelobe and main beam.   The peak power in the first
sidelobe, as a fraction of the peak of the main beam, is
$P_{n,\FS,U,I}$; the corresponding quantity for the main beam,
$P_{n,\MB,U,I}$ is unity by definition. We obtain

\begin{mathletters}
\label{robust}
\begin{equation}\label{upower}
{P_{n,\FS,U,I}} = 0.0175
\end{equation}
and the ratio of beam efficiencies as defined by equation~\ref{etabeam} is

\begin{equation}\label{etaratio} 
\left({\eta_{\FS} \over \eta_{\MB}}\right)_{U,I} = 0.0861 \; ,
\end{equation}
\end{mathletters}

\noindent Obviously, these two parameters are not independent
because, to some extent, they measure the same thing.

In the same way we can compute three {\it non-robust} parameters,

\begin{mathletters}
\label{nonrobust}
\noindent Half-power beam width:
\begin{equation}
({\it HPBW})_{U,I} = \ 0.597 {\lambda_{cm} \over \KperJy^{1/2}} \
{\rm arcmin.}
\end{equation}
Main beam efficiency:
\begin{equation}\label{etamb}
(\eta_\MB )_{U,I} = 0.840 \; ,
\end{equation}
And, combining equations~\ref{etamb} and~\ref{etaratio} to get the
total efficiency in the main beam and first sidelobe:
\begin{equation}
\left(\eta_\MB + \eta_\FS\right)_{U,I} =  0.840 (1 + 0.0861) = 0.912\; .
\end{equation}
\end{mathletters}

\noindent To calculate {\it HPBW} in equation~\ref{nonrobust}a we use
diameter $d_{\eff}= {\sqrt{4A_{\eff} \over \pi}}$ and the $HPBW$ for a 
uniformly-illuminated aperture, ${\it HPBW}=1.03 {\lambda \over
d_{\eff}}$. For the data, the observed {\it HPBW} is derived from the
Gaussian fit, and the value expected from the model  is derived from the
effective area $A_{\eff}$, which is in turn derived directly from
$\KperJy$, namely from observations of calibration sources with known
flux using a standard noise diode. This dependence on $\KperJy$ is what
makes these parameters non-robust, because $\KperJy$ is affected by
telescope  imperfections (surface roughness and subreflector
misalignment), loss in waveguide feeds, errors in source calibrator
flux, and errors in noise diode calibration temperature.

Our beam parameterization is done through Gaussian fits to the data, and
thus we should compare those fits with Gaussian fits to the ideal
uniform illumination pattern. Gaussian fits don't represent the
theoretical model perfectly, but we can compute correction factors
$P_\FS$, $E_{\FS}$, and $E_{\MB}$, and $H$ for the Gaussian fits to the
ideal uniform circular aperture. As we described above, the Gaussian-fit
derived parameters are subscripted only with ``$U$''. These correction
factors are

\begin{mathletters}
\label{cfactors}
\begin{eqnarray}
H&=&{\HPBW_U \over \HPBW_{U,I}}\\
E_{\MB}&=&{\eta_{\MB,U} \over \eta_{\MB,U,I}}\\
E_{\FS}&=&{\eta_{\FS,U} \over \eta_{\FS,U,I}}\\
P_\FS&=&\left({P_{n,\FS,U}}\over{P_{n,\FS,U,I}}\right)\; .
\end{eqnarray}
\end{mathletters}

\noindent For example, multiplication of the ideal $\HPBW_{U,I}$ by $H$
gives ${\HPBW_U}$, the value that is derived from a Gaussian fit to the
ideal beam shape; the factor 0.597 in equation \ref{nonrobust}a is
multiplied by 0.961, becoming 0.574. The ideal beam is wider at the top
and falls more rapidly than the Gaussian; this makes $H < 1$ for a
uniformly-illuminated aperture with no blocking. 

	Below, in \S\ref{blockagesection}, we will consider the effects
of blockage. There are corresponding correction factors for blockage,
which we quote here for convenience. Table~\ref{factortable} gives
values of the $(H,P,E)$ factors in equation~\ref{cfactors} for the
unblocked case (0\%) and for two different blocked cases with 10\% and
20\% blockage. Empirically, the factors change slowly with blockage, so
that these few tabular entries are sufficient.

\subsection{Comparison of Observations with the Unblocked Aperture}

        Table~\ref{unblocked} compares some of the observed parameters
with those predicted for an unblocked aperture (equations~\ref{robust}
and \ref{nonrobust}).  Arecibo's beam doesn't agree at all with these
values.  In particular, consider the two robust parameters $P_{n,\FS,U}$
and $\left({\eta_{\FS} \over \eta_{\MB}}\right)_{U}$. The observed
values are higher than the unblocked values by factors $\sim 2.2$ and
4.6, respectively.  Conventional feeds, including those used here,
produce a tapered illumination, which {\it decreases} the sidelobe level
relative to the uniformly illuminated case.  It is impossible to {\it
increase} the sidelobe level without introducing a relative decrease in
the center of the illumination pattern---a donut-like pattern.  For
Arecibo's Gregorian, this occurs naturally because of blockage of the
primary reflector by the central support structure.  We can use
Table~\ref{unblocked} to estimate the fraction of blocked area for LBW. 

\subsection{The Effects of Blockage}\label{blockagesection}

        We calculate blockage by subtracting the voltage pattern of a
circular blocking aperture from that of the illuminated aperture, taking
care to keep the power ratio equal to the area ratio. We keep the
effective area constant, because $A_{\eff}$ is what our observations
provide. We normalize the voltage at the center of the effective area's
pattern to unity and plot the above quantities versus ${\rm blocked \
area \over effective \ area}$ in Figure~\ref{blockage}. For comparing
with observations we assume that the blocked radiation is scattered into
the sky, not absorbed, so that equation~\ref{gainintegral} holds; this
reduces $\eta_{MB}$ by ${ {\rm effective \ area \over illuminated \
area}}$ relative to the assumption that the blocked radiation is
absorbed. Recall that, according to Babinet's principle, the effect of
geometrical blockage is doubled; below we will derive blockage $\sim
10\%$, which corresponds to physical blockage $\sim 5\%$.

        Figure~\ref{blockage} exhibits plots of the parameters versus
the fractional blocked area.  $X$s on these plots show the measured
quantities; the points are placed vertically using the observed value
and horizontally using the curves.  Thus, the ``derived'' blockages ${
{\rm blocked \ area \over effective \ area}}$ that correspond to the
model can be read off of the horizontal axes.  For the robust
parameters, the blockages lie somewhat below and above 0.10 in the top
two frames, respectively.  It seems reasonable to adopt a blockage of
0.10 for discussion purposes.  

	For this particular blockage, Figure~\ref{blockedbeam} exhibits
the beam shape and, also, the Gaussian fit.  The Gaussian is a
reasonably accurate rendition of the theoretical model, but has its
faults: it doesn't fall to zero between the main beam and the sidelobe
and it doesn't reproduce the details of the shape very well.  Then
again, neither the Gaussian nor the model fits the data particularly
well.  In fact, they cannot possibly do so because the sidelobe
structure changes with $\phi$, while the model is circularly symmetric. 
An additional limitation of both our parameterization and our model is
the neglect of additional sidelobes, which contribute non-trivially to
the total beam efficiency. 

	This model of a uniformly illuminated blocked aperture is useful
as an indication of expected properties of the main beam and sidelobes. 
However, it is limited and inaccurate for the following reasons:

	{\bf (1)} Most feed systems don't provide uniform illumination;
rather, the illumination is tapered.  Tapering increases $\HPBW$ and
$\eta_{\MB}$ and decreases sidelobe levels, which is opposite to
blockage.  Thus, the actual geometrical blockage must be somewhat larger
than our adopted $10\%$.  Because these two effects work in opposite
directions, we surmise that the relationships among our derived
parameters (such as $\HPBW$ and $\eta$) remain approximately correct. 

	{\bf (2)} As a telescope is pushed to higher frequencies,
imperfections in the telescope surface become increasingly important. 
If these imperfections are random, then they scatter radiation away from
the main beam according to the Ruze (1966) formulation and will decrease
$\KperJy$ and all beam efficiencies $\eta$.  Using this reduced
$\KperJy$ may make equation~\ref{nonrobust}a provide a {\HPBW} that is
too large.  The ratio of FS to MB should not be affected much unless the
correlation lengths of the surface imperfections are large---but, of
course, for most telescopes, there {\it do} exist some large scale
correlations. 

        {\bf (3)} At higher frequencies, large-scale imperfections in
the telescope surface and, also, in the mirror alignment become
important for most telescopes.  These produce significant changes in
near-in sidelobes.  These effects decrease $\KperJy$, increase $\HPBW$,
tend to increase the first sidelobe power, and tend to transfer some of
the main beam power to near-in sidelobes.  We expect $\eta_{\MB} +
\eta_{\FS}$ to decrease, but their power to go primarily into near-in
sidelobes so that the antenna temperature for an extended source will
not decrease very much.  Of course, the detailed behavior is
complicated, and in addition small-scale random surface irregularities
also become important. 

\subsection{ A More Accurate Determination of Blockage}

	Equation~\ref{nonrobust}a describes a non-robust relationship
between two robust parameters, $HPBW$ and $\lambda$; these parameters
are robust for frequencies below those where large-scale surface
irregularities affect the main beamwidth. This equation is non-robust
because it depends on \dKperJy . This particular value of \dKperJy\ is
that for a telescope with perfectly smooth reflecting surfaces and
perfect mirror alignment. This value can be calculated by the telescope
and feed engineers; we denote it by $\KperJy_{ideal}$. It can also be
measured at frequencies low enough for the telescope to be regarded as
perfect; however, such measurements rely on knowing accurate source
fluxes and cal temperatures.  

	We can observationally determine the blockage by first
rewriting equation~\ref{nonrobust}a:

\begin{equation}
\left[ \KperJy_{ideal}^{1/2} \over (0.597 \beta ) \right] = 
   { \lambda_{cm} \over HPBW_{arcmin}} \ .
\end{equation}

\noindent Here $\beta$ is a parameter that accounts for the blockage and
is given by $\beta = H \left[ HPBW \over (HPBW)_U \right] $, where $H$
comes from Table~\ref{factortable} and the $HPBW$ ratio from
Figure~\ref{blockage}.  Next we measure $HPBW$; knowing
$\KperJy_{ideal}$, we obtain the blockage factor $\beta$. 

	In practice, one measures $HPBW$ over a range of wavelengths,
using feeds with identical illumination patterns, least-squares fit for
$0.597 \beta \over \KperJy_{ideal}^{1/2}$, and calculates $\beta$.  We
obtained results at Arecibo for $\lambda = 6 \rightarrow 70$ cm; the fit
is very good and, assuming $\KperJy_{ideal} = 10.5$, it yields $\beta =
0.88$.  From Figure~\ref{blockage}, this corresponds to about $13\%$
blockage.  We believe that the blockage so derived should be accurate,
but it does depend on knowing a reliable value for $\KperJy_{ideal}$. 

\section{ {\boldmath $\phi$} STRUCTURE IN THE FIRST SIDELOBE} \label{fs}

        We introduced the Fourier decomposition of the first sidelobe in
section~\ref{fsintro}; we obtained a series with 8 points and we use
$N=6$ to reconstruct the sidelobe.  This allows us to reconstruct four
Fourier components, numbered zero to three.  Figure~\ref{fsi} exhibits
the $\az$ dependence of these components, and we briefly discuss these
data to illustrate the power of this representation, because the
$(az,za)$ dependences of these Fourier coefficients reveal their
probable production mechanisms. 

        The average amplitude (the zeroth Fourier component, top panel)
seems to increase slightly near transit, which probably reflects the
slightly increased blockage that occurs there.

        The angles of maximum response $\phi_{max}$ of the first and
third Fourier components exhibit a striking dependence with \az: the
points follow the slope of the dotted line, meaning that these
components are fixed with respect to the ground.  The third Fourier
component would respond to blockage by the triangle structure, which is
a large, major structural entity that supports the feed structure; its
{\az} dependence implies that it is doing just that.  The interpretation
for the first Fourier component is less obvious. 

        The position angles of the second Fourier component are
independent of \az, implying that they are fixed with respect to the
azimuth arm. This is consistent with the blockage of the azimuth arm,
which is a linear structure roughly centered on the illumination pattern
and should produce a second Fourier component.

	The first and third Fourier components are fixed with respect to
the ground, in contrast to the behavior of the coma lobe, which is fixed
with respect to the azimuth arm. We conclude that the sidelobe structure
and coma lobe are, for the most part, unrelated. We reaffirm this
conclusion in \S \ref{fsquv}, where we discuss the sidelobe
polarization.

\section{ THE POLARIZED STOKES PARAMETERS} \label{otherstokes}

	One derives the beam structure of the polarized Stokes
parameters by observing an unpolarized source, or by subtracting off the
polarization of a polarized source to simulate an unpolarized source. An
unpolarized source contains power in all polarizations, so beam
structure in the polarized Stokes parameters reflects enhanced gain of
one polarization over its orthogonal counterpart.  Thus, in this section
we are discussing the response of the telescope in Stokes $Q$, $U$, and
$V$ to an unpolarized source. Note that this response, unlike the Stokes
$I$ response, can be positive or negative. We assume that the on-axis
polarization has been corrected as described in the companion paper
Heiles et al (2001), so that $(Q,U,V) = 0$ at beam center. 

\subsection{Squint and Squash: Parameters for the Polarized Main Beam}

	For each polarized Stokes parameter such as $Q$, we express the
response of the main beam with the first two terms of a two-dimensional
Taylor expansion.  The first derivative term, conventionally called
``beam squint'', is odd and corresponds to displacement of the beam
center in the two orthogonal polarizations. Beam squint is characterized
by an angular magnitude and a direction and has a two-lobed structure on
the sky, with one positive and one negative lobe. The second term, which
we call ``beam squash'', is even and corresponds to a difference in beam
width between the two polarizations. Beam squash is characterized by a
magnitude and an orientation and has a four-lobed structure, with lobes
on opposite side of beam center having identical signs. We assume that
the coma lobe is unpolarized.

        To be more precise, consider Stokes $Q$ as an example, which is
polarization $X$ minus $Y$. We define the normalized power pattern for a
polarization $X$ as $G_X \over G_{max}$.\footnote{ Recall that above we
defined the unsubscripted $G$ (which includes $G_{max}$) to apply to ${
{\rm Stokes \ I} \over 2}$.} Then the power received from the source in
each polarization is

\begin{mathletters}
\label{X1}   
\begin{equation}
P_{X} = S_X \exp  \left[ -
{\left( \theta + {\delta \theta_Q \over 2} \right)^2 
\left( 1 - 
\alpha_{\coma} {\theta_{\coma} 
\over \Theta_0} 
\right) \over \left( \Theta + {\delta \Theta_Q \over 2}\right) ^2} 
\right]
\end{equation}
\begin{equation}
P_{Y} = S_Y \exp \left[ -
{\left( \theta - {\delta \theta_Q \over 2} \right)^2 
\left( 1 - 
\alpha_{\coma} {\theta_{\coma} 
\over \Theta_0} 
\right) \over \left( \Theta - {\delta \Theta_Q \over 2}\right) ^2} 
\right] \; .
\end{equation}
\end{mathletters}

\noindent where $S_X$ and $S_Y$ are the source fluxes in the two
polarizations. Recall that $\Theta$, $\theta_Q$, and $\Theta_Q$ all
depend on  $\phi$. The unsubscripted $P$ is the sum (Stokes $I$), while
the difference gives the measured Stokes $Q$:

\begin{equation} \label{qlsfit}
P_{Q} = Q P_n + {I \over 2} \left[ 
        {\partial P_n \over \partial \theta_Q} \delta \theta_Q (\phi) +
        {\partial P_n \over \partial \Theta_Q} \delta \Theta_Q (\phi)
        \right] \; .
\end{equation}

\noindent The quantities $\delta \theta_Q (\phi)$ and $\delta \Theta_Q
(\phi)$ are not themselves the beam squint and squash. Rather, the
squint and squash are represented by  (magnitude, position angle), 
$(\theta_{squint}, \phi_{squint})$ and $(\theta_{squash},
\phi_{squash})$, and are given by

\begin{mathletters}
\begin{equation} 
\delta \theta_Q = \theta_{squint} \cos( \phi - \phi_{squint})
\end{equation}
\begin{equation} 
\delta \Theta_Q = \Theta_{squash} \cos 2( \phi - \phi_{squash}) \; .
\end{equation}
\end{mathletters}

\noindent The quantity $\delta \theta_Q$ varies as $\cos \phi$ because
it is like a pointing offset and has a {\it direction}. In contrast, the
 quantity $\delta \Theta_Q$ varies as $\cos 2 \phi$: in beam squash, the
beamwidth varies as $\cos 2 \phi$ so it only has an {\it orientation}.

\subsection{ Results for Beam Squint and Squash}

        We least-squares fit equation~\ref{qlsfit} to the data,
regarding $P_n$ as known. This is acceptable, even though $P_n$ is
itself a result of a least-squares fit, because its data are Stokes $I$
with high S/N. We do this in all three polarized Stokes parameters.
Figures \ref{mbq} and \ref{mbu} show representative plots for Stokes
$(Q,U)$. The squint and squash amplitudes are in the few arcsecond range
and there is significant variation, but no easily interpretable physical
dependence, on $(\za,\az)$. On the other hand, Stokes $V$
(Figure~\ref{mbv}) is more smoothly behaved: the squint is somewhat
smaller and maintains a constant $\phi$, while the squash is very small.

	For the linearly polarized Stokes parameters $(Q,U)$, we expect
beam squash but no squint (Tinbergen 1996).  The prediction regarding
squints is not satisfied at all.  Instead, beam squints for Stokes
$(Q,U)$ are comparable to that for $V$.  The behaviors of
$\phi_{squint}$ for $(Q,U)$ are uninterpretable in simple terms and
provide no hint about the cause of the squint.  The beam squash
amplitude variations are also uninterpretable.  The angles
$\phi_{squash}$ should be fixed with respect to the azimuth arm; very
roughly speaking, they are.  However, they should differ by $45^\circ$
between $Q$ and $U$; they don't. 

	For Stokes $V$, one expects beam squint but no beam squash 
(see discussion by Troland and Heiles 1982). This prediction is
reasonably well satisfied. The squint amplitude is roughly constant and
its angle $\phi_{squint}$ is constant, fixed with respect to the azimuth
arm as it should be. The squash is small, far smaller than for Stokes
$(Q,U)$, although it is not zero.

	The reasons for the severe departure of Stokes $(Q,U)$ from
theory are completely unclear to us. The absence of a clearly defined
dependence with on $az$ suggests a combination of effects that are fixed
with respect to the ground and the az arm; the obvious candidates are 
aperture blockage and secondary/tertiary mirror misalignment.

\subsection{Fourier Parameters for the Polarized Sidelobes} \label{fsquv}

	For each polarized Stokes parameter such as $Q$, we follow \S
\ref{fs} and express the response of the sidelobe in terms of Fourier
components. Figures \ref{fsq}, \ref{fsu}, and \ref{fsv} display these
coefficients for the three polarized Stokes parameters. The zeroth
Fourier coefficient describes the net polarization averaged over the
sidelobe ring and is shown in the top panels. The first sidelobe is
strongly linearly polarized. 

	In Stokes $Q$ and $U$, the first Fourier component dominates and
sometimes produces fractional polarizations ${(Q^2 + U^2)^{1/2} \over I}
\gtrsim 0.5$. Their angles of maximun response $\phi_{max}$ have a
complicated behavior: their slopes with respect to $az$ indicate that
they are approximately fixed with respect to the ground, but
$\phi_{max}$ suffers a $180^\circ$ jump near $az = (-20^\circ,
20^\circ)$ for $(Q, U)$ respectively. The dependences for the second and
third Fourier components are even less clear. 

	Because of the dominance of the first Fourier component, the
sidelobe linear polarization has a strong component that is fixed with
respect to the ground (apart from the $180^\circ$ jumps), in contrast to
the behavior of the coma lobe, which is fixed with respect to the
azimuth arm. Thus we reaffirm the conclusion of \S \ref{fsi}: the
sidelobe structure and coma lobe are, for the most part, unrelated.

	The angles shown in Figures \ref{fsq}, \ref{fsu}, and \ref{fsv}
are not the position angles of linear polarization. Rather, they are the
angles $\phi$, defined in Figure \ref{ibeameg}, at which the particular
Fourier coefficient is maximum. In contrast, Figure \ref{fspa} exhibits
the position angle of linear polarization $P\!A_{linpol} = 0.5
\tan^{-1}{U \over Q}$ for the four Fourier coefficients. As expected,
$P\!A_{linpol}$ for the zeroth component follows the azimuth arm in its
rotation  with respect to the sky. For the other components the behavior
is more complicated, which occurs because $\phi_{max}$ changes with
azimuth: in other words, the $\phi$ of maximum response moves around the
sidelobe ring differently in $Q$ and $U$ as the azimuth changes, which
leads to rotation of $P\!A_{linpol}$ with respect to the azimuth arm as
it rotates when tracking a source.

	In Stokes $V$, the first Fourier component dominates and
produces fractional circular polarization ${V \over I} \sim 0.1$. For
the first and third Fourier components the angles of maximun response
$\phi_{max}$ are nearly constant, being fixed with respect to the
ground. The dependence for the second Fourier components is unclear,
perhaps because it is so weak. 

\subsection{ The Effects of Beam Squint and Squash on Extended Sources}

        Suppose one is observing a large-scale feature where the
brightness temperature $T_B$ varies with position.  One can express this
variation by a two-dimensional Taylor expansion.  The first derivative
is described by a single term with a magnitude and direction; the second
derivative is described by three terms, ${d^2 T_B \over d\az^2}$, ${d^2
T_B \over d\za^2}$, ${d^2 T_B \over d\az d\za}$, each of which is an
ordinary second derivative with a direction.  {\bf In this section
and its figures, $T_B$ refers to Stokes $I$ so is twice its
conventionally used value}. 

	Beam squint, by its nature, responds to the first derivative and
only slightly to the second; beam squash responds primarily to the
second.  To illustrate the effects of polarized beam structure, we
present plots of the response of the polarized beam to these
derivatives.  In these plots we adopt derivatives of 1 K arcmin$^{-1}$
and 1 K arcmin$^{-2}$ to facilitate scaling to other measured
derivatives, not because these values are necessarily realistic; also,
in these figures the brightness temperatures refer to Stokes $I$, not $I
\over 2$, so they are twice the conventionally used values.  The
response comes from three separable causes: main beam squint, main beam
squash, and the first sidelobe polarization structure.  We show these
responses individually and, also, we show the total response. 

        Figures~\ref{sqq} and \ref{squ} show Stokes $Q$ and $U$. In
these two figures, the position angles $P\!A$ are defined {\it on the sky
with the conventional astronomical definition}. For the first-derivative
case, the dominant contribution comes from the first sidelobe, which
contributes far more than the main beam squint. The amplitudes are
roughly constant with $az$. The angles of maximum response are, {\it
very} roughly, fixed on the sky (they change less rapidly than $az$),
which is consistent with the first sidelobe's effects being caused by
blockage. For the second derivative case, the first sidelobe
contribution dominates over main beam squash, but not by much. 

        Figure~\ref{sqv} shows the Stokes $V$ result. In this figure,
$P\!A$ is the angle of maximum response {\it defined relative to the
azimuth arm.} The total amplitudes for both derivatives are smaller by a
factor $\sim 10$ than for Stokes $(Q,U)$.  For the first derivative, the
$P\!A$ is independent of $\az$, as expected from the squint result. The
sidelobe is again the dominant contribution, but not overwhelmingly so:
in fact, its amplitude is about twice that of the main beam squint but
the direction is opposite. This has the curious effect of making the
{\it amplitude} of the total beam response about the same with and
without the sidelobe---but reversing the {\it direction} of gradient for
maximum response! For the second derivative, the first sidelobe and main
beam squash contribute comparably and the $P\!A$ behavior is
correspondingly complicated.

        We emphasize that these particular results can produce serious
systematic errors when measuring the polarization of extended emission. 
The results represented in Figures~\ref{sqq}, \ref{squ}, \ref{sqv} are
contributions in the {\it polarized} Stokes parameters $(Q,U,V)$ that
arise from spatial gradients in the {\it total intensity} Stokes
parameter $I$.  The fractional polarization of extended emission tends
to be small, so spatial gradients in $I$ can be very serious. 
Correcting for them at Arecibo is a complicated business because of the
$P\!A$ variation with $(\az,\za)$.  It is also an uncertain business,
especially for $(Q,U)$ and somewhat less so for $V$, because these
variations are unpredictable and must be determined empirically.
Presumably, corrections at conventional telescopes are more
straightforward. 

        In that spirit, these plots are more useful as guides of what
can be reliably measured at Arecibo than they are for correction of the
measurements.  Consider, for example, measuring Zeeman splitting of the
21-cm line in emission, which involves measuring Stokes $V$.
Figure~\ref{sqv} shows that a brightness temperature gradient ${dT_B
\over d\theta} = 1$ K arcmin$^{-1}$ produces an artificial temperature  
$V_{\rm art}\sim 0.04$ K response in Stokes $V$; let this ratio be
denoted by ${\cal K} \equiv {V_{\rm art} \over dT_B/d\theta} \sim 0.04$
arcmin. Then it's easy to show that, for a Land\'e $g$ factor 2.8 Hz
$\mu$G$^{-1}$, the corresponding artifical magnetic field is $B_{\rm
art} = 1690 {\cal K} {dv \over d\theta}$ $\mu$G, where ${dv \over   
d\theta}$ is the velocity gradient in km s$^{-1}$ arcmin$^{-1}$ and
$\cal K$ is in arcmin.  For a representative ${dv \over d\theta} = 1$ km
s$^{-1}$ deg$^{-1}$ we get $B_{\rm art} \sim 1.1$ $\mu$G.  Heiles (1996)
presents statistics of the velocity gradient measured with a 36 arcmin
$HPBW$; gradients of 1 km s$^{-1}$ deg$^{-1}$ are not uncommon.
Gradients might be larger when measured with smaller $HPBW$.  Typical
values of $B$ are in the $\mu$G range, so this effect can be---but is
perhaps not always---serious!

\section{ IMAGES OF THE BEAM IN ALL STOKES PARAMETERS} \label{greyfigs}

        Expressing the beam properties in terms of parameters is 
especially appropriate for quantitative work. However, for getting an
intuitive feel for the situation there's nothing like an image. So here
we provide images for one of our several hundred beam maps. As with our
previous detailed figures, these images are all for LBW (1175 MHz)
observing B1749+096; the source was setting near Arecibo's $za$ limit of
$20^\circ$, with $(\az,\za) \approx (64^\circ, 18.4^\circ)$.

        The main beam parameters are as follows:

         For Stokes $I$, $(\Theta_0, \Theta_1) = (4.00, 0.36)$ arcmin
(both are HPBW); $\phi_\beam= 91.1^\circ$; $\alpha_{\coma} = 0.048$;
$\phi_{\coma}= 41.4^\circ$; mean height of first sidelobe= 0.029.

        For Stokes $Q$, $(\theta_{squint}, \phi_{squint}) = (0.019',
-173^\circ)$ and $(\Theta_{squash}, \phi_{squash}) = (0.11', 19^\circ)$.
As in Figure~\ref{mbq}, $Q$ is oriented $45^\circ$ to the azimuth arm.

        For Stokes $U$, $(\theta_{squint}, \phi_{squint}) = (0.056',
-42^\circ)$ and $(\Theta_{squash}, \phi_{squash}) = (0.061',
-47^\circ)$. As in Figure~\ref{mbu}, $U$ is aligned with the azimuth
arm.

        For Stokes $V$, $(\theta_{squint}, \phi_{squint}) = (0.045',
-10^\circ)$ and $(\Theta_{squash}, \phi_{squash}) = (0.007',
37^\circ)$. 

        The beam is considerably elliptical, which is a result of the
high zenith angle. The Stokes $Q$ beam shows the classic four-lobed   
pattern, beam squash, for linear polarization; in contrast, the $U$ beam
is polluted with a large unpredicted squint and a huge linearly   
polarized sidelobe.  Theoretically, the beam squashes in $(Q,U)$ must be
aligned with their orientations, and in particular the two four-lobed  
patterns should be rotated $45^\circ$ with respect to each other; this
is not the case! The $V$ beam is reasonably well-behaved. We commented
on these aspects in more detail above.

\acknowledgements

        It is a pleasure to acknowledge insightful and helpful comments
by the referee, which substantially improved several aspects of this
paper. This work was supported in part by NSF grant 95-30590 to CH.

\clearpage

\section{FIGURE CAPTIONS}

\figcaption{Grey-scale/contour image of a main beam having
$\alpha_{\coma} = 0.2$, $\phi_{\coma} = 22.5^\circ$, $\Theta_0 =
3.4$ arcmin, $\Theta_1 = 1$ arcmin (both $\Theta$'s are \HPBW~in
this figure), and $\phi_\beam = -67.5^\circ$. The dashed lines
intersect the true beam center; the dashed-dot lines intersect the
assumed (from the telescope pointing) beam center, and are the paths
of the telescope in the observing pattern. Solid contours are $(0.1,
0.2, \dots)$ of the peak; dashed contours are $(0.01, 0.02,
\dots)$.\label{ibeameg} }

\figcaption{Blockage-sensitive, measurable parameters versus the ratio
of blocked area to effective area.  Solid lines show the parameters as
derived by fitting the main beam and first sidelobe to Gaussians; dotted
lines show the actual parameters for the theoretical model. The top two
frames show the robust parameters of equation \ref{robust} and the third
and fourth show the non-robust parameters of equation \ref{nonrobust}.
\label{blockage} } 

\figcaption{Normalized power pattern $P_n$, and also the integrand
$\theta P_n$, for {\bf (1)} the standard model of a uniformly
illuminated  blocked aperture (solid, dash-dot lines), and {\bf (2)} its
Gaussian-fit counterpart (dash, dash-dot-dot-dot lines). The squares and
diamonds are representative data points for $P_n$ from the LBW feed at
1415 MHz, obtained by averaging different cuts in one single observing
pattern. \label{blockedbeam} }

\figcaption{$\az$ dependence of the first four Fourier components of the
first sidelobe in Stokes $I$ for LBW at 1175 MHz.  The dotted lines
correspond to a component being fixed with respect to the ground.  Solid
lines are amplitude and the points are position angle $\phi_{max}$,
which is the angle $\phi$ at which the component peaks ($\phi$ is
defined in Figure~\ref{ibeameg}).  For the $n$th Fourier component ($n =
0 \rightarrow 3$), each point occurs $n$ times over each $360^\circ$
interval in $\phi$, and all of those possibilities are plotted to better
clarify the {\az} dependence.  \label{fsi} } 

\figcaption{$(\za,\az)$ dependence of amplitudes (in arcmin) and 
$(\phi_{squint}, \phi_{squash})$ of beam squint and squash for Stokes
$Q$. (Here $\phi$ is labeled ``PA''). The dotted line corresponds to
being fixed with respect to the sky or ground. {\bf In this figure,
{\boldmath $Q$} is oriented {\boldmath $45^\circ$} to the azimuth arm
and fixed with respect to it.} \label{mbq} }  

\figcaption{$(\za,\az)$ dependence of amplitudes (in arcmin) and 
$(\phi_{squint}, \phi_{squash})$ of beam squint and squash for Stokes
$U$. (Here $\phi$ is labeled ``PA''). The dotted line corresponds to
being fixed with respect to the sky or ground. {\bf In this figure,
{\boldmath $U$} is aligned with the azimuth arm and fixed with respect
to it.} \label{mbu} }  

\figcaption{$(\za,\az)$ dependence of amplitudes (in arcmin) and 
$(\phi_{squint}, \phi_{squash})$ of beam squint and squash for Stokes
$V$. (Here $\phi$ is labeled ``PA''). The dotted line corresponds to
being fixed with respect to the sky or ground. \label{mbv} } 

\figcaption{$\az$ dependence of the first four Fourier components of the
first sidelobe in Stokes $Q$ for LBW at 1175 MHz.  The dotted lines
correspond to a component being fixed with respect to the ground.  Solid
lines are amplitude and the points are position angle $\phi_{max}$,
which is the angle $\phi$ at which the component peaks ($\phi$ is
defined in Figure~\ref{ibeameg}).  For the $n$th Fourier component  ($n
= 0 \rightarrow 3$), each point occurs $n$ times over each $360^\circ$
interval in $\phi$, and all of those possibilities are plotted to better
clarify the {\az} dependence.  \label{fsq} } 

\figcaption{$\az$ dependence of the first four Fourier components of the
first sidelobe in Stokes $U$ for LBW at 1175 MHz.  The dotted lines
correspond to a component being fixed with respect to the ground.  Solid
lines are amplitude and the points are position angle $\phi_{max}$,
which is the angle $\phi$ at which the component peaks ($\phi$ is
defined in Figure~\ref{ibeameg}).  For the $n$th Fourier component ($n
= 0 \rightarrow 3$), each point occurs $n$ times over each $360^\circ$
interval in $\phi$, and all of those possibilities are plotted to better
clarify the {\az} dependence.  \label{fsu} } 

\figcaption{$\az$ dependence of the position angle of linear
polarization $P\!A_{linpol}$ for the first four Fourier components of
the first sidelobe for LBW at 1175 MHz ($P\!A$ is the conventional
astronomical definition).  The dotted lines correspond approximately to
$P\!A_{linpol}$ being fixed with respect to the azimuth arm. 
\label{fspa} } 

\figcaption{$\az$ dependence of the first four Fourier components of the
first sidelobe in Stokes $V$ for LBW at 1175 MHz.  The dotted lines
correspond to a component being fixed with respect to the ground.  Solid
lines are amplitude and the points are position angle $\phi_{max}$,
which is the angle $\phi$ at which the component peaks ($\phi$ is
defined in Figure~\ref{ibeameg}).  For the $n$th Fourier component ($n =
0 \rightarrow 3)$, each point occurs $n$ times over each $360^\circ$
interval in $\phi$, and all of those possibilities are plotted to better
clarify the {\az} dependence.  \label{fsv} }

\figcaption{$\az$ dependence of the telescope's response to the first
spatial derivative in brightness temperature (top two panels) and to the
second derivative (bottom two panels), all for Stokes $Q$. $P\!A$ is
direction of maximum response. {\bf In this figure, {\boldmath $Q$} and
{\boldmath $P\!A$} are defined on the sky with the conventional
astronomical definition.} $T_B$ refers to Stokes $I$ so is twice is
conventionally used value. \label{sqq} } 

\figcaption{$\az$ dependence of the telescope's response to the first
spatial derivative in brightness temperature (top two panels) and to the
second derivative (bottom two panels), all for Stokes $U$. $P\!A$ is
direction of maximum response. {\bf In this figure, {\boldmath $U$} and
{\boldmath $P\!A$} are defined on the sky with the conventional
astronomical definition.} $T_B$ refers to Stokes $I$ so is twice is
conventionally used value. \label{squ} }

\figcaption{$\az$ dependence of the telescope's response to the first
spatial derivative in brightness temperature (top two panels) and to the
second derivative (bottom two panels), all for Stokes $V$. $P\!A$ is the
$\phi$ of maximum response, {\bf defined relative to the azimuth arm as in
Figure~\ref{ibeameg}}. $T_B$ refers to Stokes $I$ so is twice is
conventionally used value. \label{sqv}} 

\figcaption{Grey-scale/contour of the Stokes $I$ main beam and first
sidelobe reconstructed from the derived parameters. Solid contours are
$(0.1, 0.2, \dots)$ of the peak; dashed contours are $(0.01, 0.02,
\dots)$. \label{mkfig1i} } 

\figcaption{Grey-scale/contour of the Stokes $Q$ main beam and first
sidelobe reconstructed from the derived parameters. Black contours are
for areas with negative $Q$ with the grey scale tending towards  white;
white contours are positive $Q$ with the greyscale tending towards
black. Contours are in percent of Stokes $I$ at beam center and spaced
by $0.4\%$; the $0\%$ contour is omitted. As in Figure~\ref{mbq}, $Q$ is
oriented $45^\circ$ to the azimuth arm.\label{mkfig1q} } 

\figcaption{Grey-scale/contour of the Stokes $U$ main beam and first
sidelobe reconstructed from the derived parameters. Black contours are
for areas with negative $U$ with the grey scale tending towards  white;
white contours are positive $U$ with the greyscale tending towards
black. Contours are in percent of Stokes $I$ at beam center and spaced
by $0.4\%$; the $0\%$ contour is omitted. As in Figure~\ref{mbu}, $U$ is
aligned with the azimuth arm.\label{mkfig1u} } 

\figcaption{Grey-scale/contour of the Stokes $V$ main beam and first
sidelobe reconstructed from the derived parameters. Black contours are
for areas with negative $V$ with the grey scale tending towards  white;
white contours are positive $V$ with the greyscale tending towards
black. Contours are in percent of Stokes $I$ at beam center and spaced
by $0.2\%$; the $0\%$ contour is omitted. \label{mkfig1v} } 

\clearpage

\begin{deluxetable} {cccccccccc} 
\footnotesize
\tablecaption{ REPRESENTATIVE OBSERVED PARAMETER VALUES \label{repdata}}
\tablehead{
\colhead{RCVR} & \colhead{FREQ} & \colhead{SOURCE} & \colhead{\za} & 
\colhead{$P_{n,\FS}$} & \colhead{${\eta_{\FS} \over \eta_{\MB}}$} &
\colhead{$\KperJy$} & \colhead{{\it HPBW}} & \colhead{$\eta_\MB$} & 
\colhead{$\eta_\MB + \eta_\FS$} 
}
\startdata
430G  &  430 & B1634+269 & 10 & .039 & 0.33 & 10.3 & 10.9 & 0.66 & 0.88 \nl
\nl
LBW   & 1175 & B1634+269 & 10 & .043 & 0.32 & 8.7  & 4.0 & 0.56 & 0.75 \nl
LBW   & 1415 & B1634+269 & 10 & .046 & 0.32 & 7.5  & 3.4 & 0.50 & 0.67 \nl
LBW   & 1666 & B1634+269 & 10 & .042 & 0.31 & 6.9  & 2.9 & 0.48 & 0.62 \nl
\enddata
\tablecomments{{\za} is in degrees and \HPBW~is in arcmin.} 
\end{deluxetable}

\clearpage

\begin{deluxetable} {lccc} 
\footnotesize
\tablecaption{FACTORS IN EQUATIONS~\ref{cfactors} VERSUS 
BLOCKAGE \label{factortable}}
\tablewidth{250pt}
\tablehead{
\colhead{FACTOR}& \colhead{$0\%$} & \colhead{$10\%$} & \colhead{$20\%$}
}
\startdata
$H$         &  0.961   & 0.955      & 0.952      \\ 
$P_{\FS}$    &  1.038   & 1.038      & 1.033      \\ 
$E_{\MB}$   &  1.060   & 1.067      & 1.070      \\ 
$E_\FS$    &  0.865   & 0.925      & 0.947      \\ 
\enddata
\end{deluxetable}

\clearpage

\begin{deluxetable} {cccccccccc} 
\footnotesize
\tablecaption{COMPARISON OF OBSERVED AND UNBLOCKED MODEL PARAMETERS 
\label{unblocked}}
\tablehead{
\colhead{RCVR} & \colhead{FREQ} & \colhead{${P_{{\FS}} \over (P_{\FS})_U}$} & 
\colhead{${(\eta_{\FS} / \eta_{\MB}) \over (\eta_{\FS} / \eta_{\MB})_U}$} &
\colhead{$d_{\eff}$} & \colhead{${{\HPBW} \over ({\HPBW})_U}$} & 
\colhead{${\eta_{\MB} \over (\eta_{\MB})_U} $} & 
\colhead{$\eta_{\MB} + \eta_{\FS} \over (\eta_{\MB} + \eta_{\FS})_U$}
}
\startdata
430G  &  430 &  2.2 & 4.7  & 190  & 0.87 & 0.74 & 0.92 \nl
\nl
LBW   & 1175 &  2.4 & 4.6 & 175  & 0.80 & 0.63 & 0.79 \nl
LBW   & 1415 &  2.5 & 4.6 & 162  & 0.75 & 0.56 & 0.70 \nl
LBW   & 1666 &  2.3 & 4.4 & 156  & 0.71 & 0.54 & 0.65 \nl
\enddata
\tablecomments{See \S\ref{unifill} for parameter definitions. $d_{\eff}$
is in meters.}
\end{deluxetable}

\clearpage

\begin{figure}[h!] 
\plotone{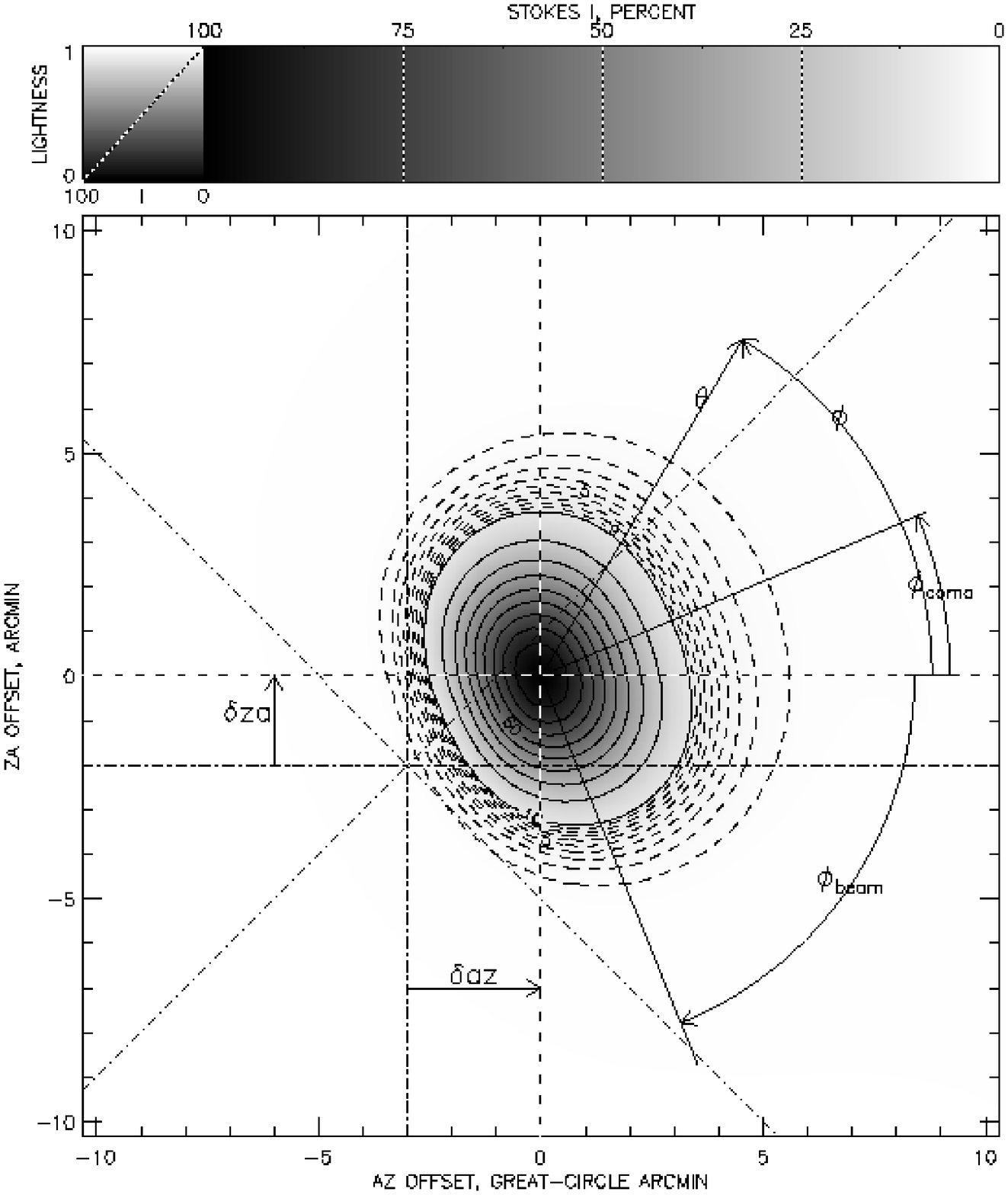}
\end{figure}

\clearpage

\begin{figure}[h!]
\plotone{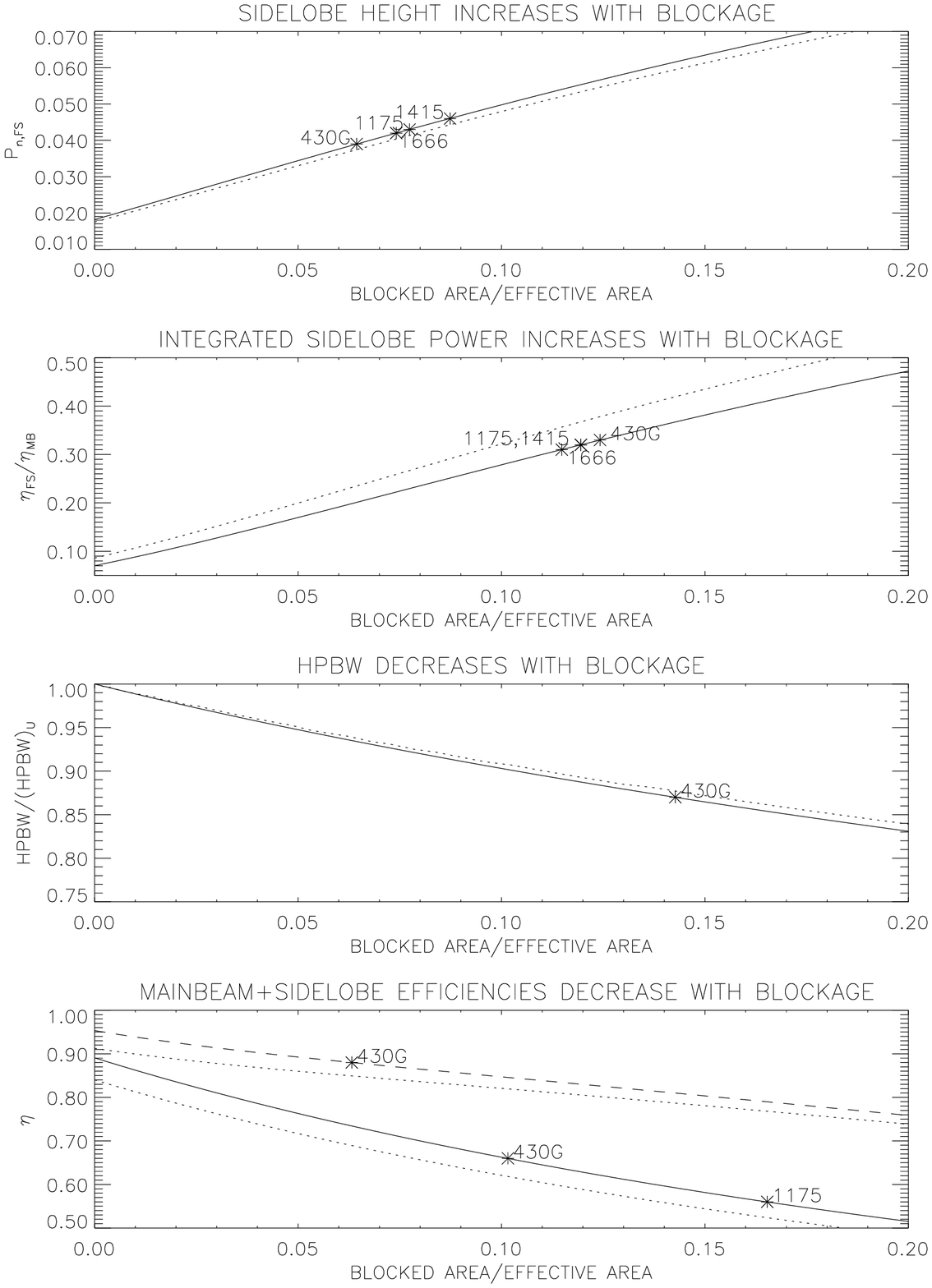}
\end{figure}

\clearpage

\begin{figure}[h!]
\plotone{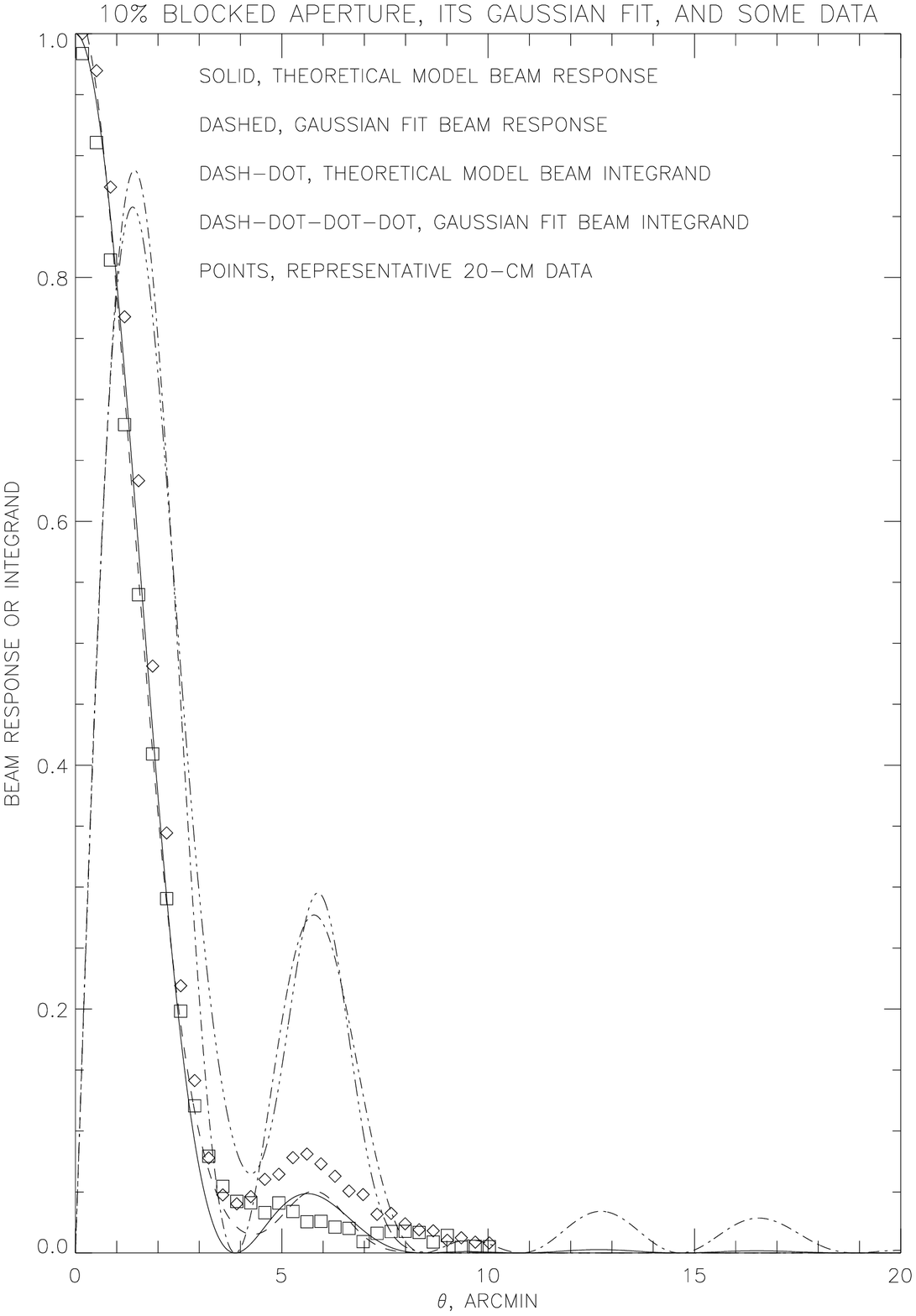}
\end{figure}

\clearpage

\begin{figure}[h!]
\plotone{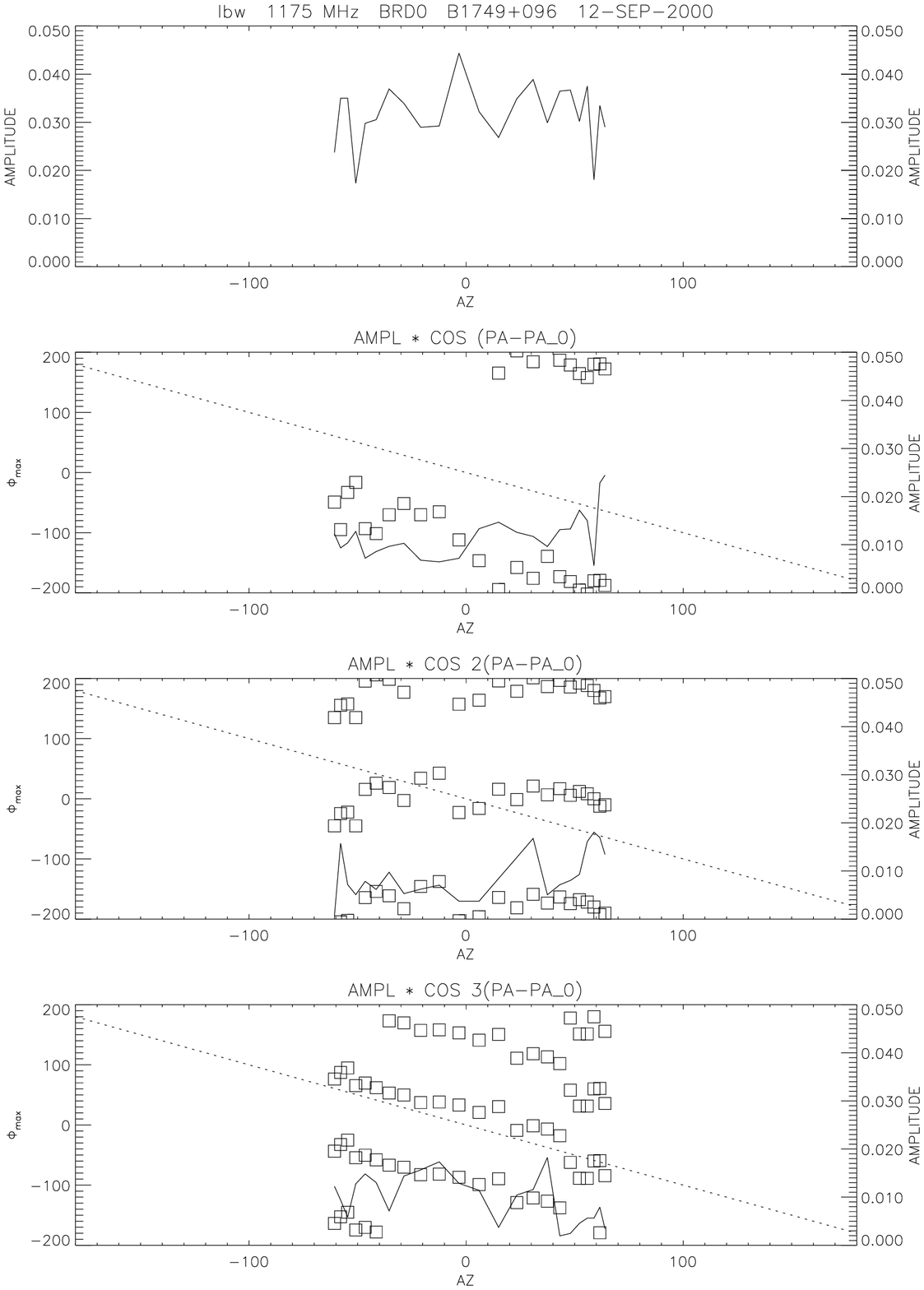}
\end{figure}

\clearpage

\begin{figure}[h!]
\plotone{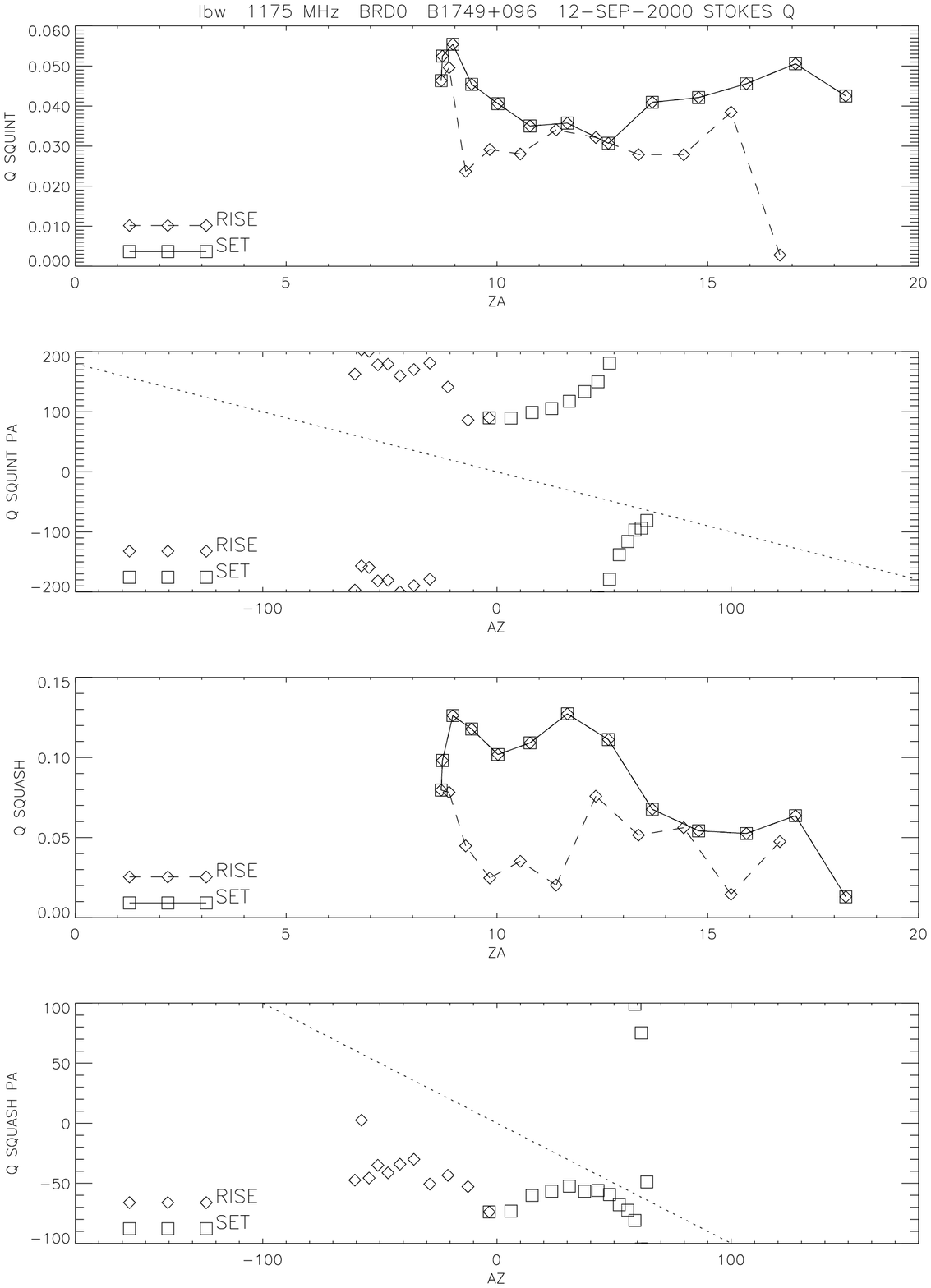}
\end{figure}

\clearpage

\begin{figure}[h!]
\plotone{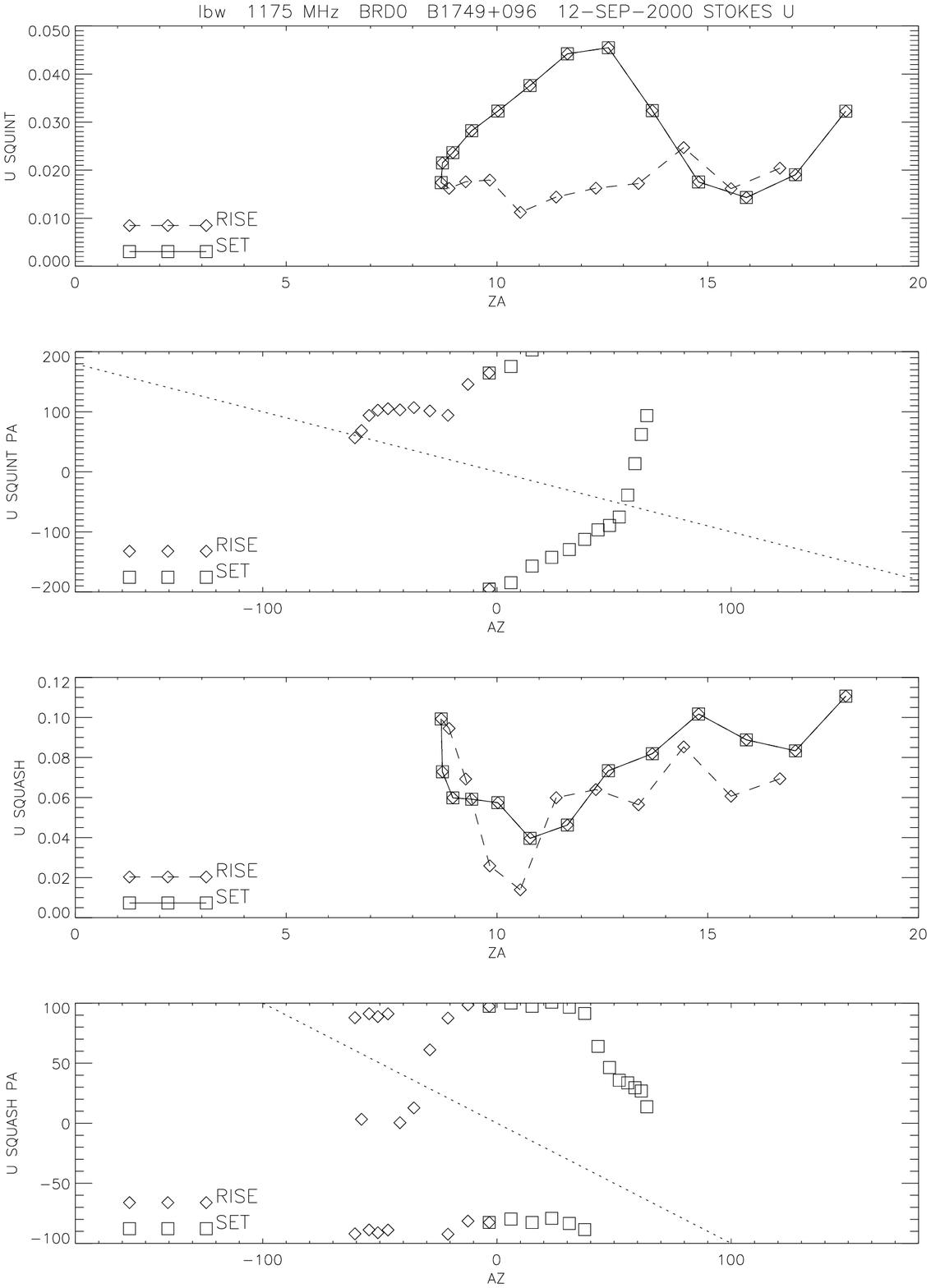}
\end{figure}

\clearpage

\begin{figure}[h!]
\plotone{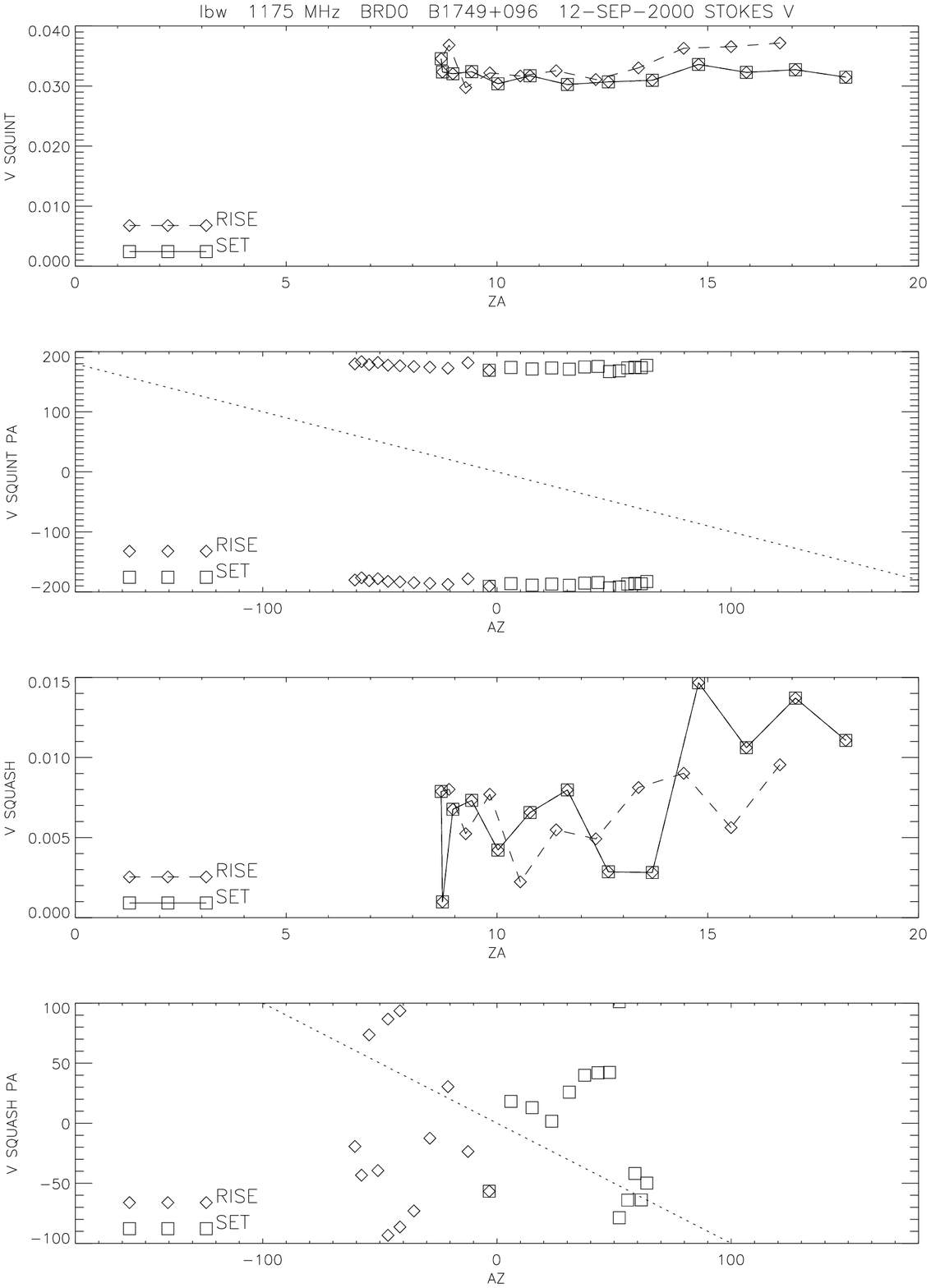}
\end{figure}


\clearpage

\begin{figure}[h!]
\plotone{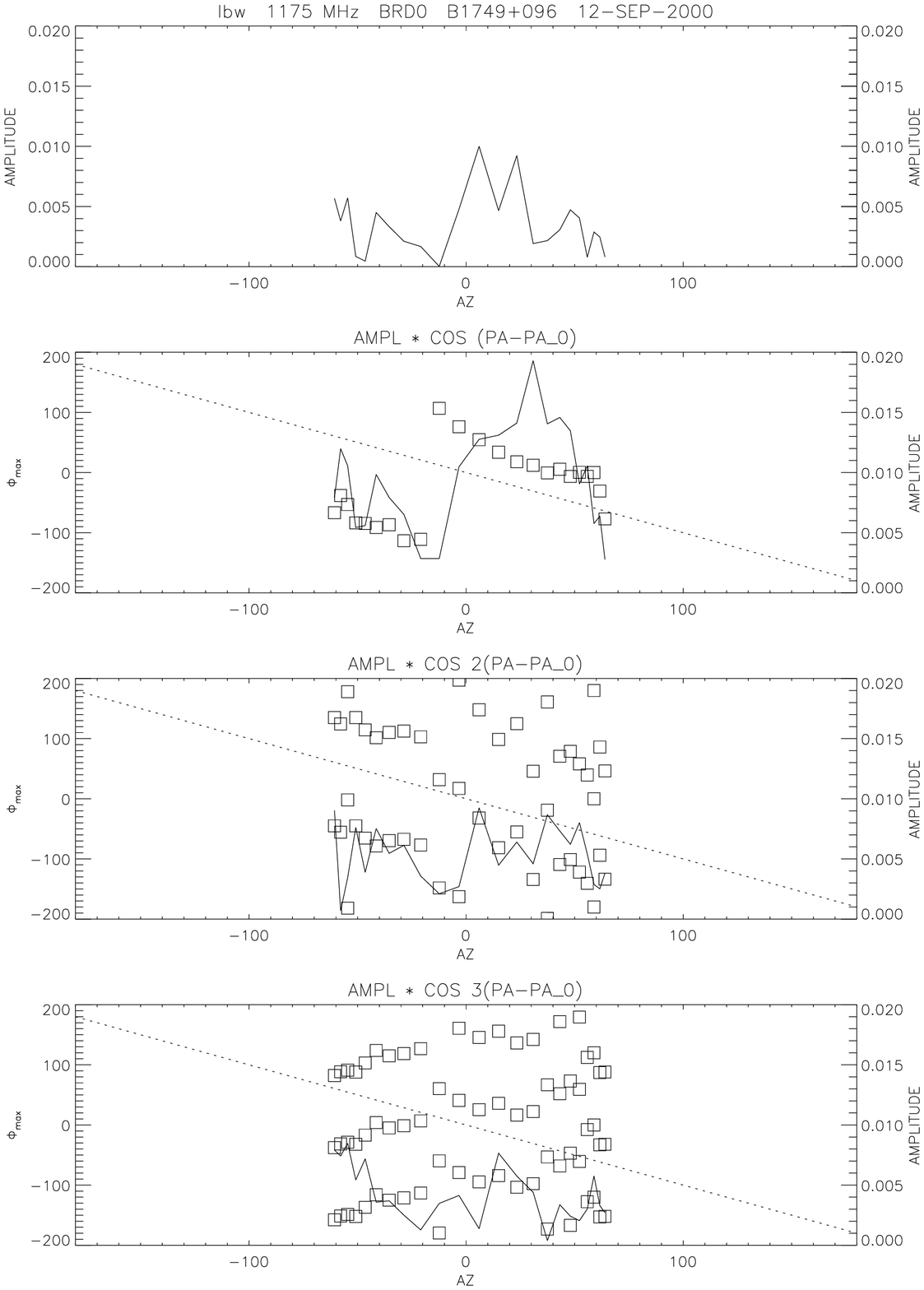}
\end{figure}

\clearpage

\begin{figure}[h!]
\plotone{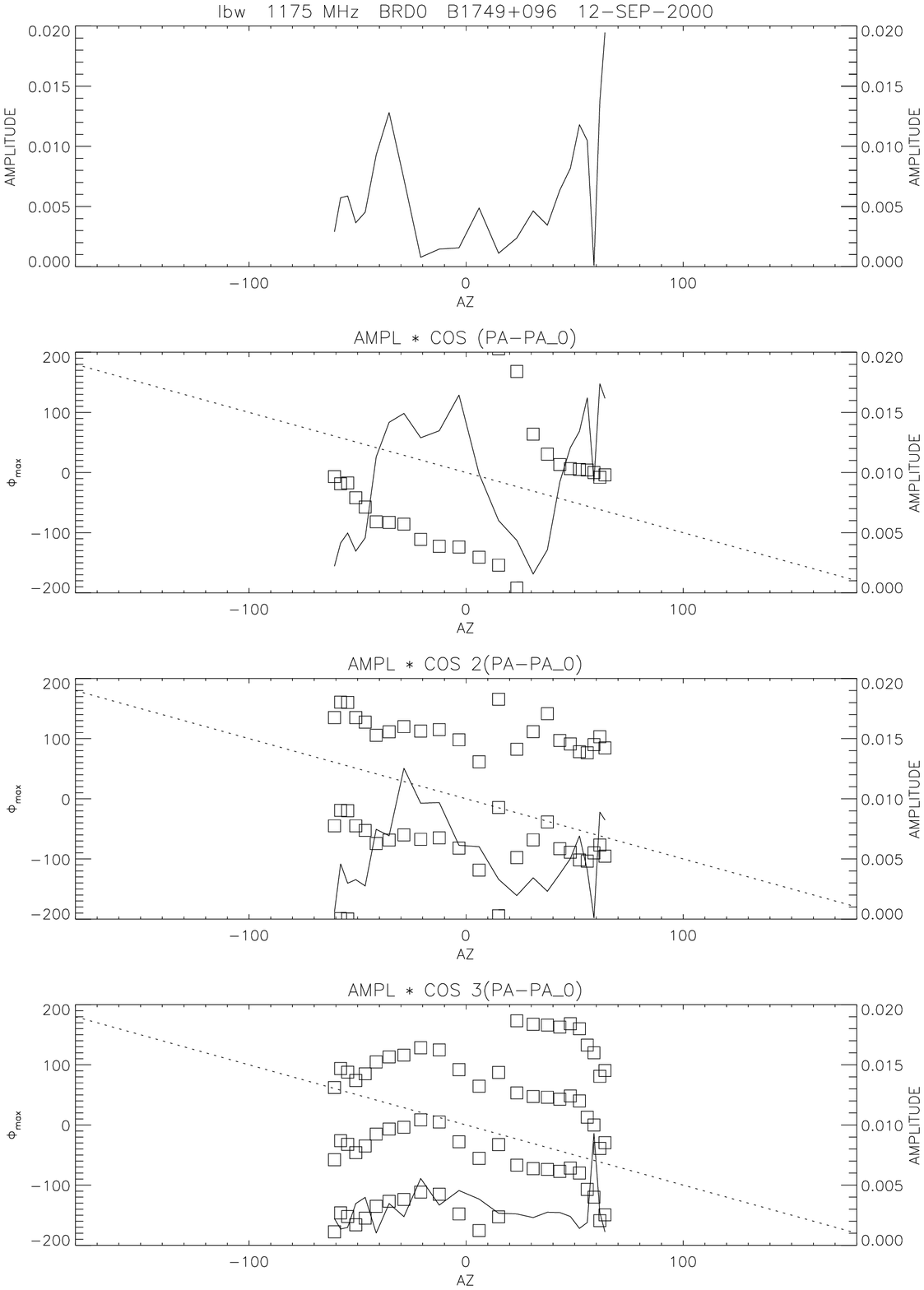}
\end{figure}

\clearpage

\begin{figure}[h!]
\plotone{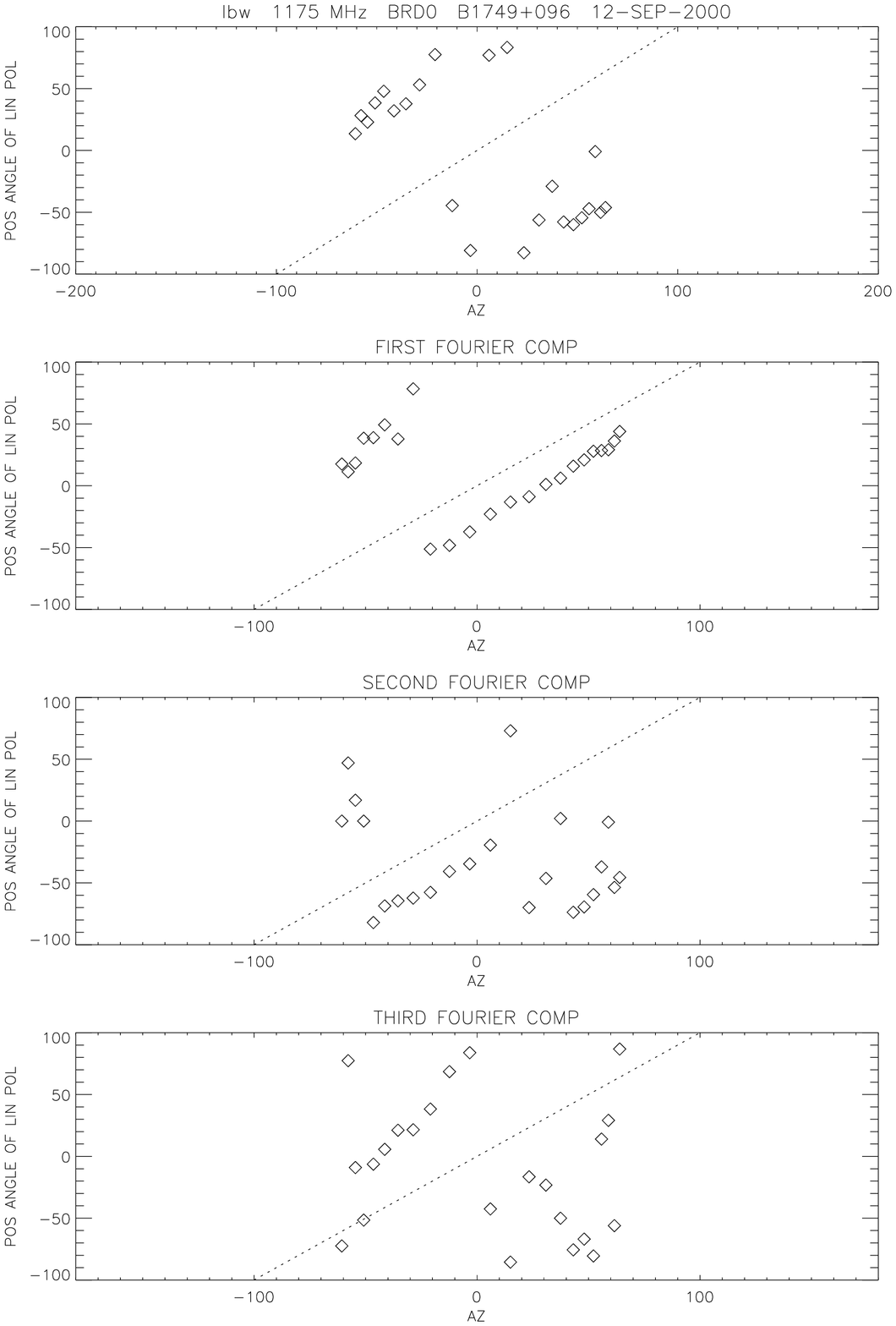}
\end{figure}

\clearpage

\begin{figure}[h!]
\plotone{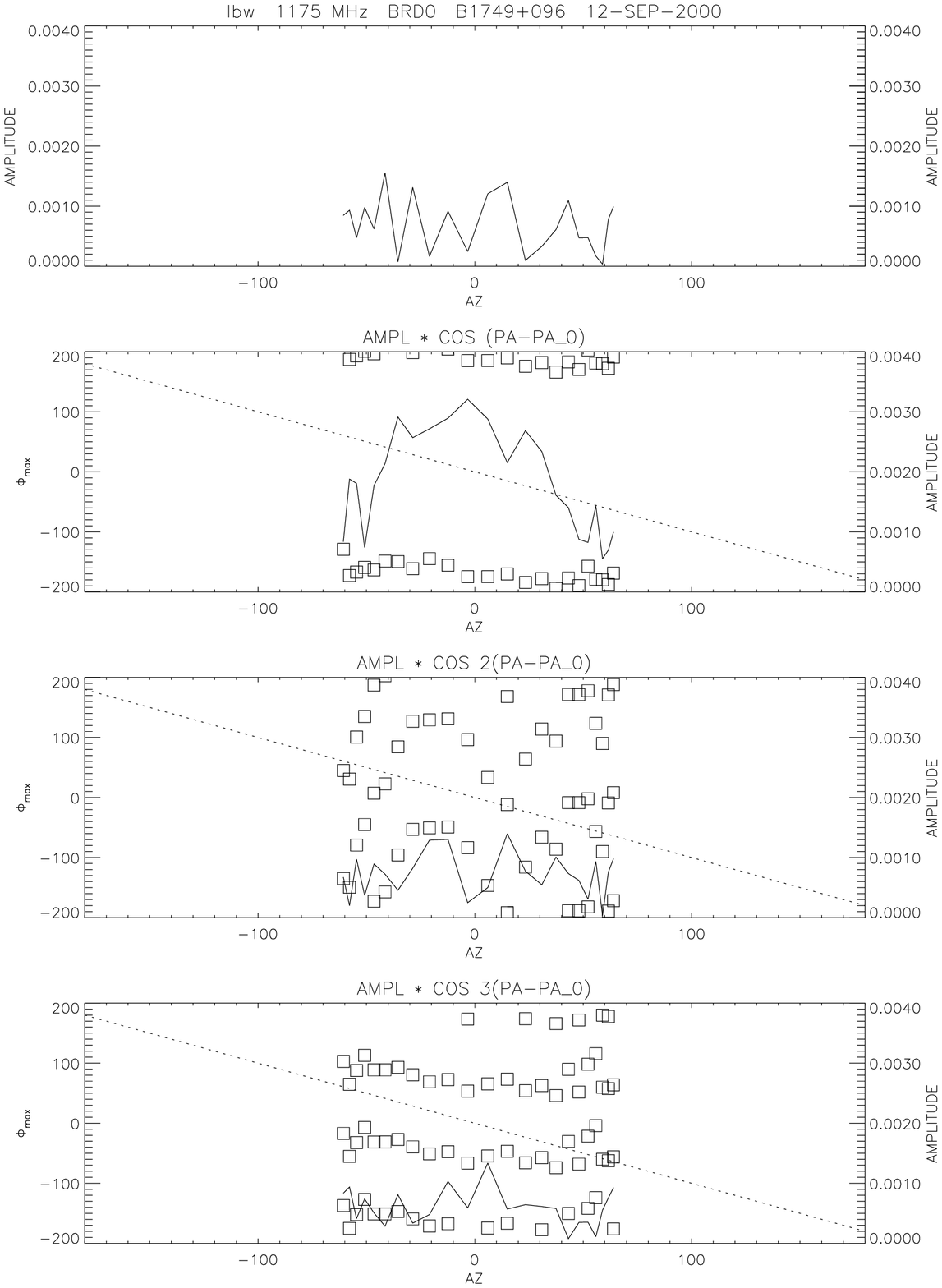}
\end{figure}

\clearpage

\begin{figure}[h!]
\plotone{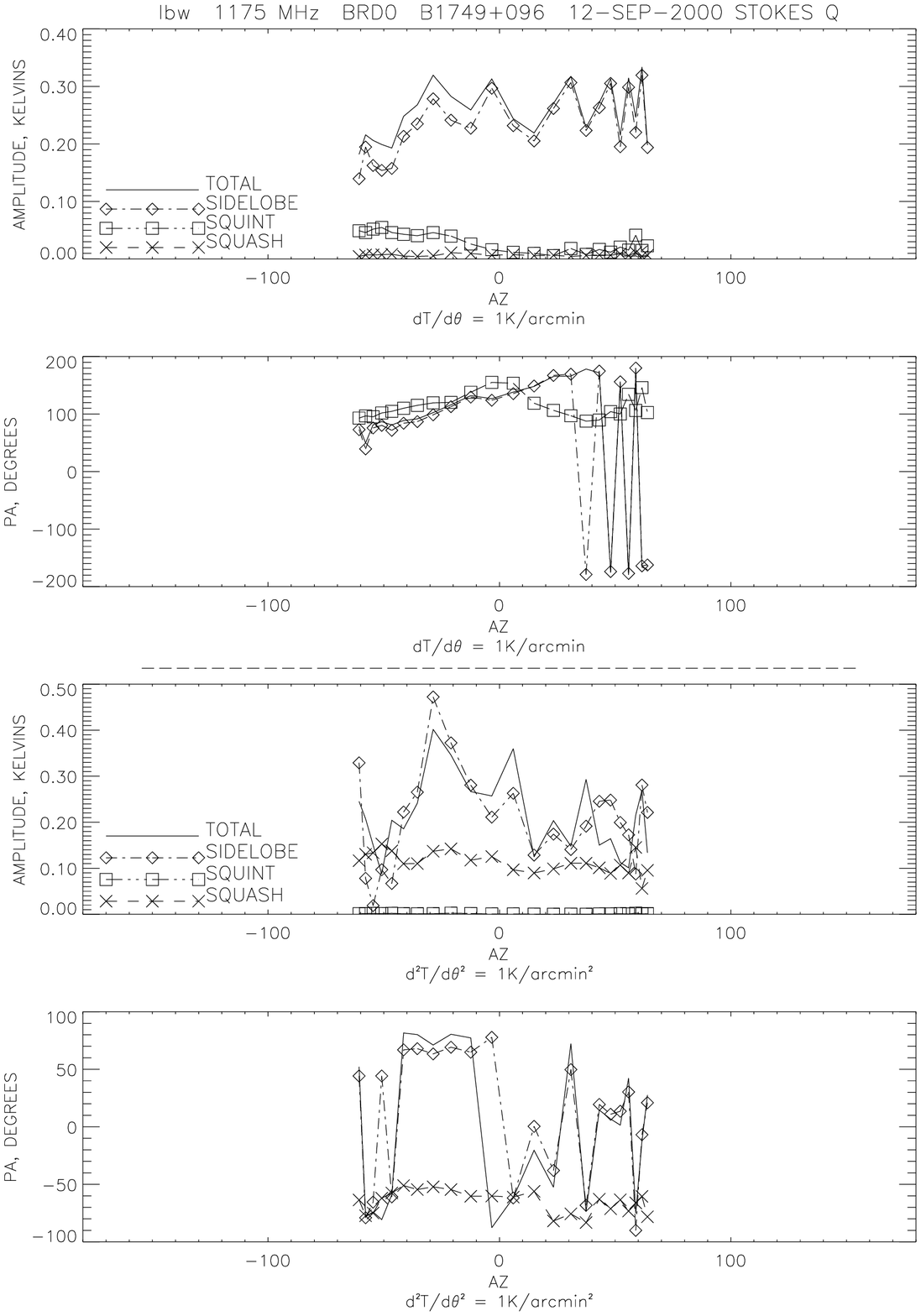}
\end{figure}

\clearpage

\begin{figure}[h!]
\plotone{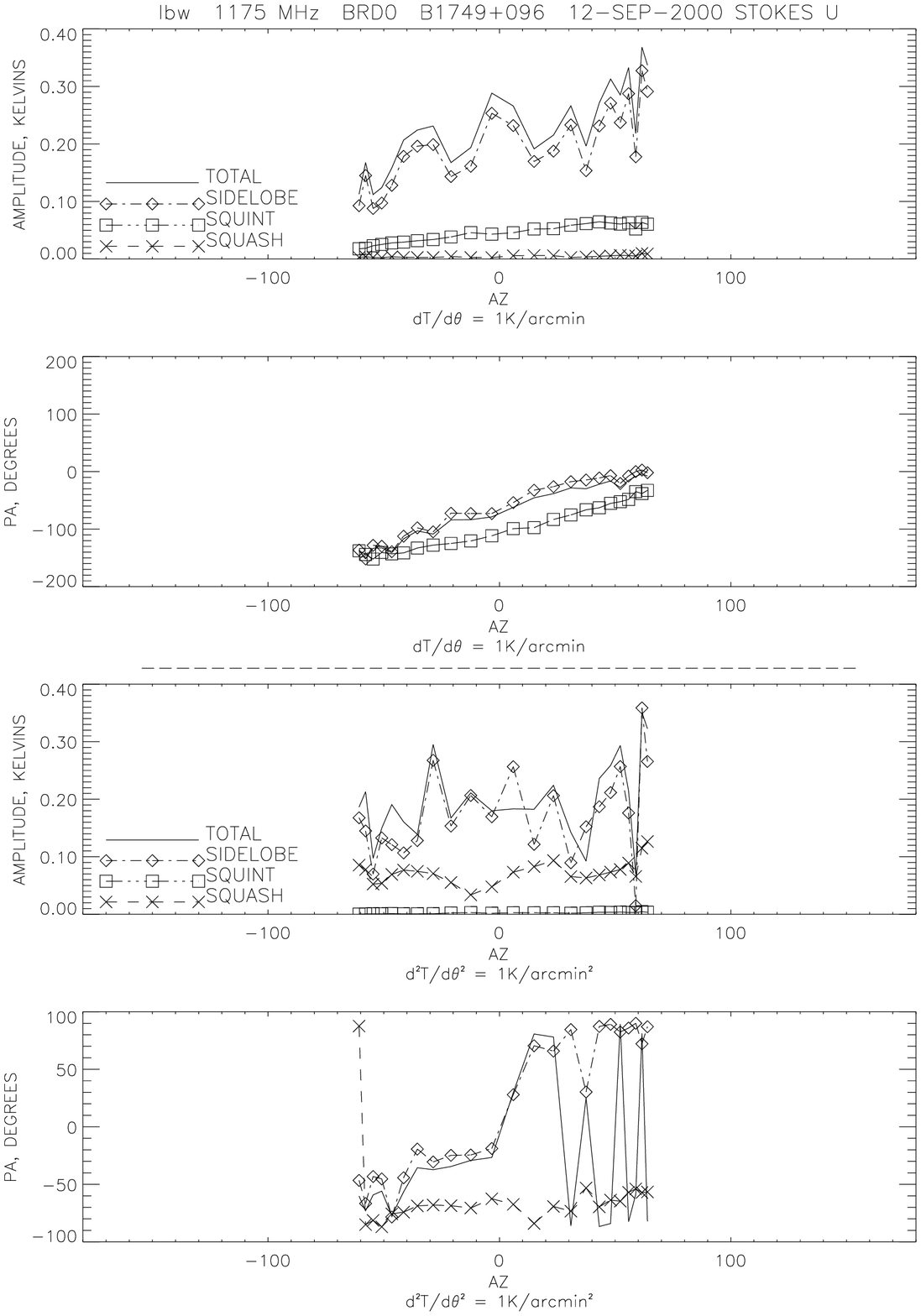}
\end{figure}

\clearpage

\begin{figure}[h!]
\plotone{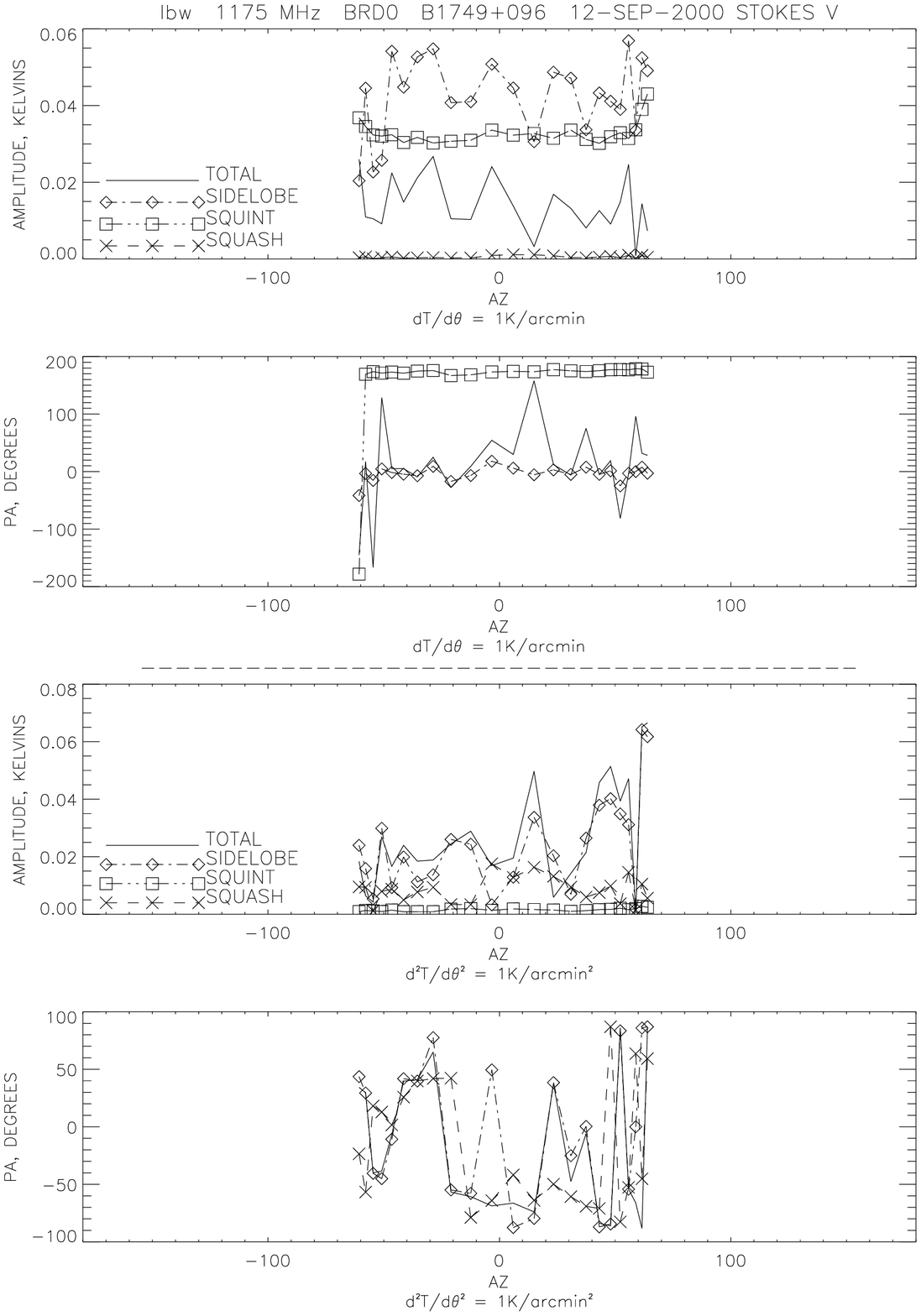}
\end{figure}

\begin{figure}[h!]
\plotone{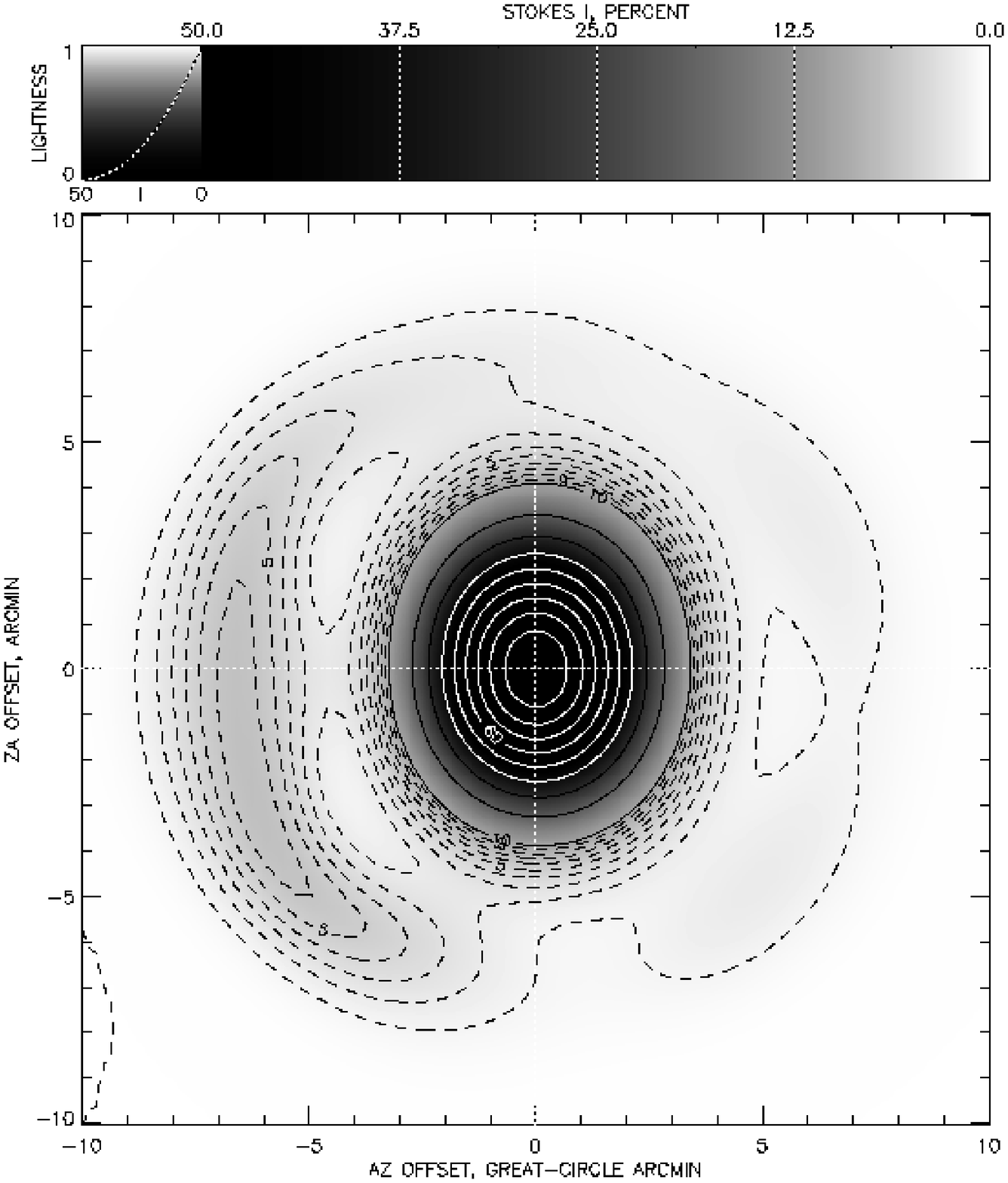}
\end{figure}

\begin{figure}[h!]
\plotone{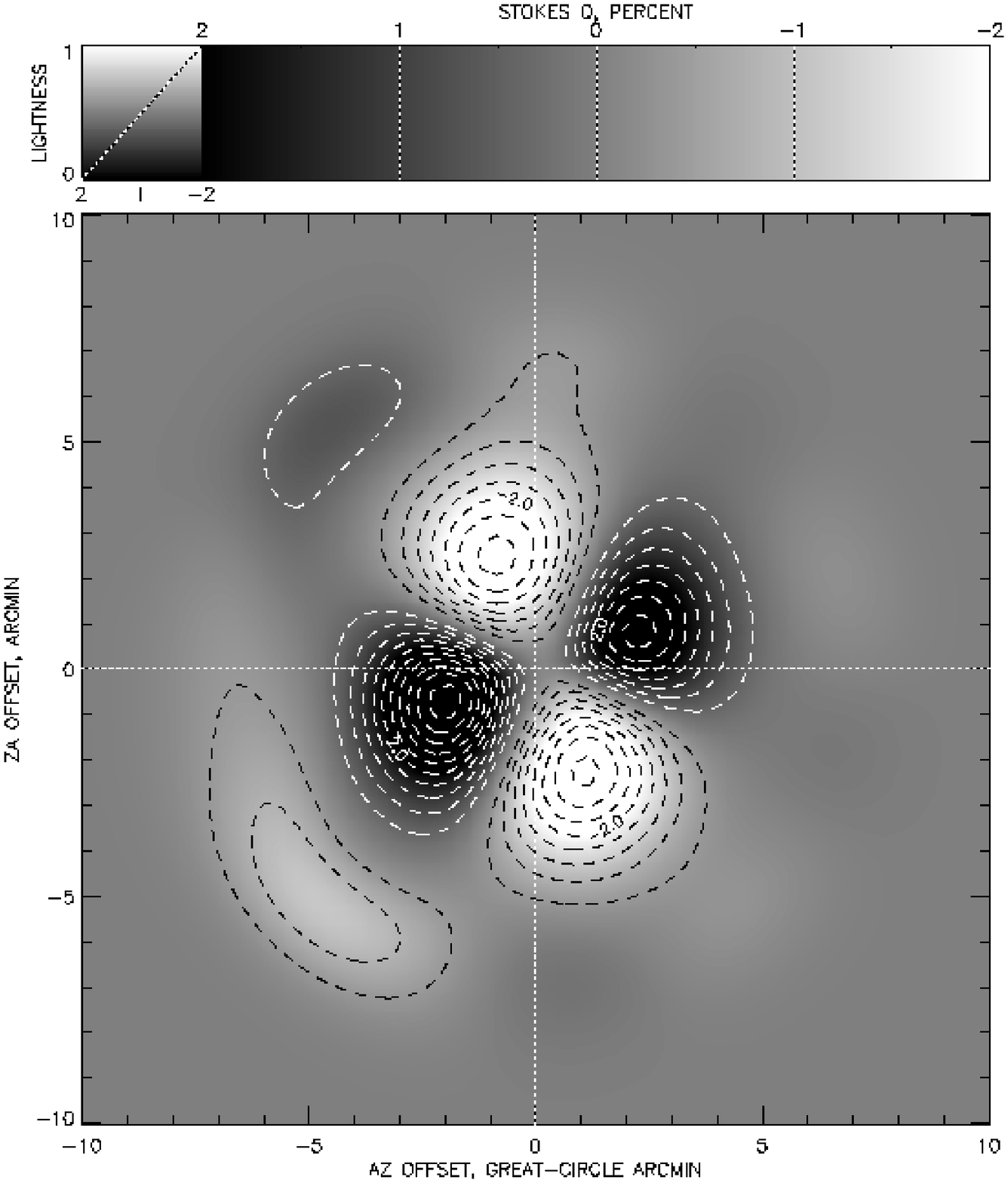}
\end{figure}

\begin{figure}[h!]
\plotone{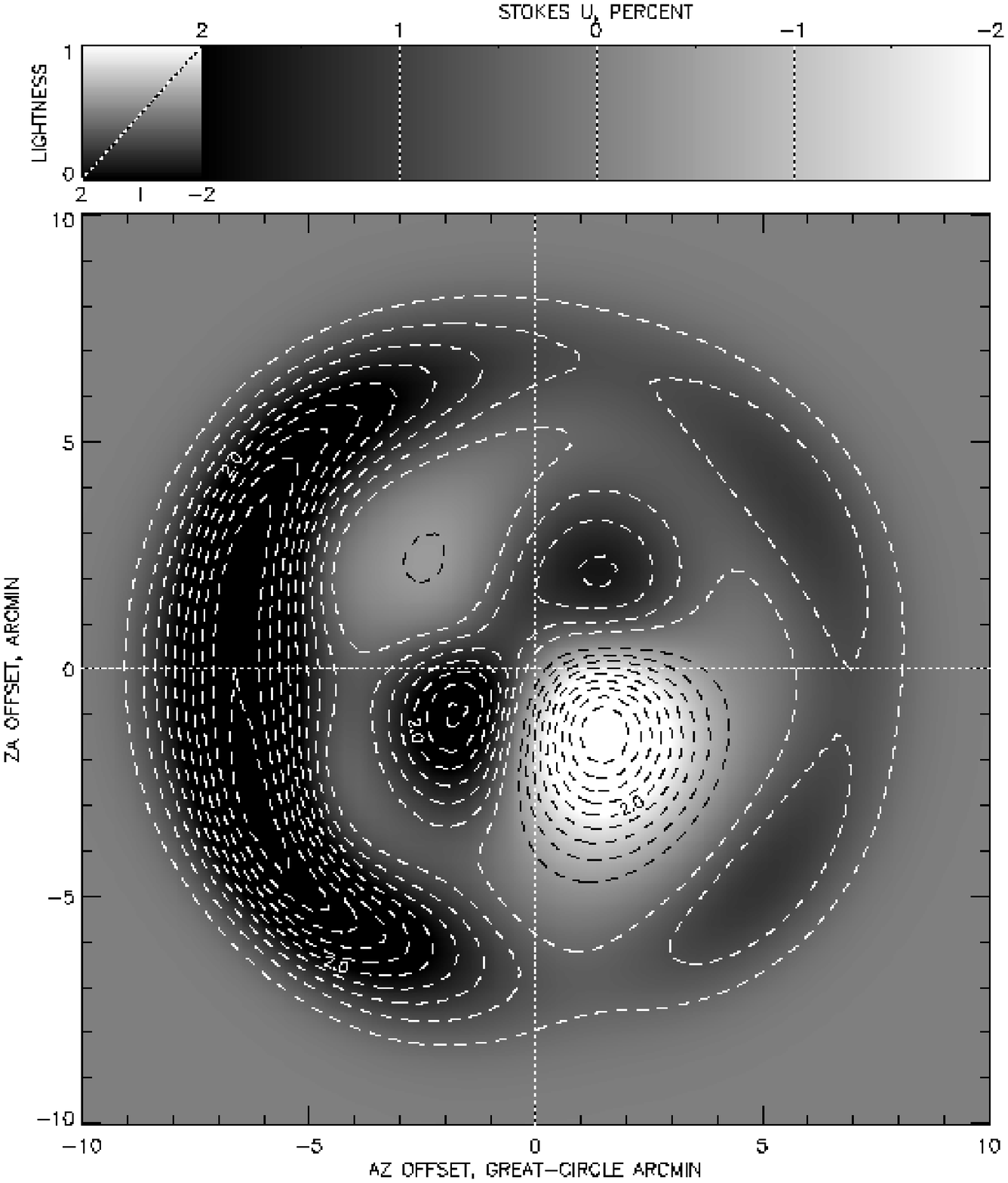}
\end{figure}

\begin{figure}[h!]
\plotone{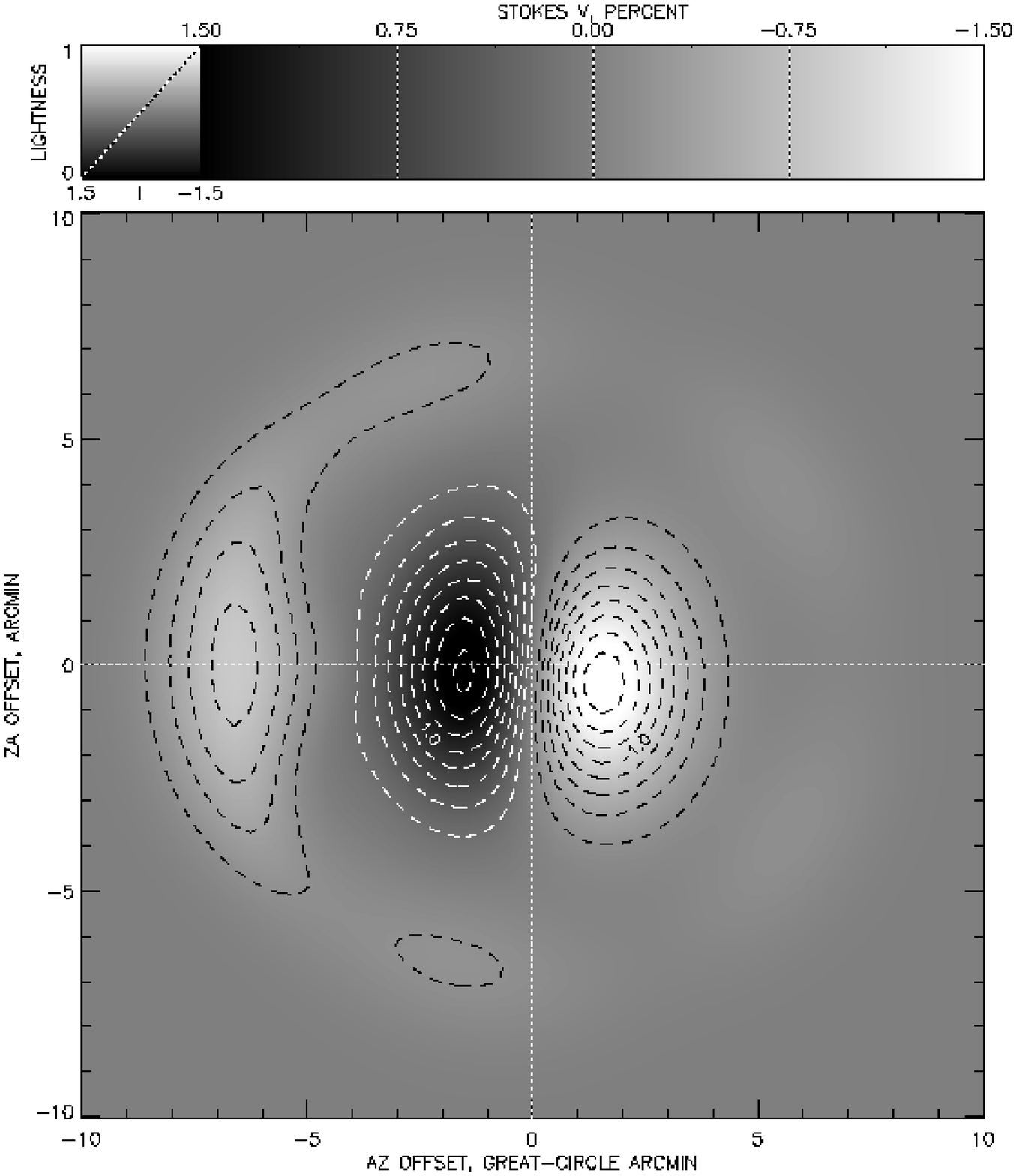}
\end{figure}

\enddocument
\end